\documentclass[twocolumn]{svjour3}          
\smartqed  
\usepackage{graphicx}
\usepackage{times}

\usepackage[tbtags]{amsmath}
\usepackage{color}

\usepackage{mylatexcommands}

\newcommand {\figdir} {./figures}
\newcommand {\ttd}{\mbox{\small T}} 

\journalname{CEAS Space Journal}
\begin{document}
\rmfamily
\title{Exploratory numerical experiments with a macroscopic theory of interfacial interactions
}


\author{D.  Giordano \and  P.  Solano-L\'{opez} \and J. M. Donoso}


\institute{D. Giordano \at
                  ESA - Estec, Keplerlaan 1, 2201 AZ Noordwijk, The Netherlands \\
                  Tel.: +31 71 565 4392 \\
                 \email{Domenico.Giordano@esa.int}           
                 \and
                 P. Solano-L\'{opez} ,  J. M. Donoso\at Dpt de Fisica Aplicada, ETSIAE, Universidad Polit\'{e}cnica de Madrid, Plaza Cardenal Cisneros 3, 29040 Madrid, Spain \\
                 \email{Pablo.Solano@upm.es, Josemanuel.Donoso@upm.es} }

\date{Received: date / Accepted: date}

\maketitle

\begin{abstract}
Phenomenological theories of interfacial interactions are founded on the core idea to model macroscopically the thin layer that forms between media in contact as a two-dimensional continuum (surface phase or interface) characterised by physical properties per unit area; the temporal evolution of the latter is governed by surface balance equations whose set acts as bridging channel in between the governing equations of the volume phases. 
These theories have targeted terrestrial applications since long time and their exploitation has inspired our research programme to build up, on the same core idea, a macroscopic theory of gas-surface interactions targeting the complex phenomenology of hypersonic reentry flows as alternative to standard methods in aerothermodynamics based on accommodation coefficients.  
The objective of this paper is the description of methods employed and results achieved in the exploratory study that kicked off our research programme, that is,
the unsteady heat transfer between two solids in contact in planar and cylindrical configurations with and without interface. 
It is a simple numerical-demonstrator test case designed to facilitate quick numerical calculations but, at the same time, to bring forth already sufficiently meaningful aspects relevant to thermal protection due to the formation of the interface.
The paper begins with a brief introduction on the subject matter and a review of relevant literature within an aerothermodynamics perspective.
Then the case is considered in which the interface is absent. 
The importance of tension (force per unit area) continuity as boundary condition on the same footing of heat-flux continuity is recognised and the role of the former in governing the establishment of the temperature-difference distribution over the separation surface is explicitly shown.
Evidence is given that the standard temperature-continuity boundary condition is just a particular case.
Subsequently the case in which the interface is formed between the solids is analysed.
The coupling among the heat-transfer equations applicable in the solids and the balance equation for the surface thermodynamic energy more conveniently formulated in terms of the surface temperature is discussed.
Results are illustrated and commented for planar and cylindrical configuration; they show unequivocally that the thermal-protection action of the interface turns out to be driven exclusively by thermophysical properties of the solids and of the interface; accommodation coefficients are not needed.
Future work of more fluid-dynamics nature is mentioned in the concluding section.
  
\end{abstract}

\section{Introduction} \label{intro}

The physics housed inside the transition layer between two different media and the micro/macroscopic phenomena occurring therefrom have always attracted interest and attention of scientists concerned with a wide variety of applications across numerous departments of scientific knowledge. 
Historically, capillarity phenomena came first under scrutiny; 
since the times of Da Vinci, recognised \cite[(note 1, page 551)]{cw1857apc} first investigator, they were studied thoroughly by reputed scientists of the calibre of Young \cite{ty1805ptrs}, Laplace \cite{psl1805tmc}, Gauss \cite{cfg1877w5}, Poisson \cite{sdp1831}, Gibbs \cite{jg1876tca,jg1878tca,jg1878ajs}, Maxwell \cite{jcm1890sp2}, van der Waals \cite{jvdw1895ansen}, Poincar\'{e} \cite{hp1895c}, Einstein \cite{ae1901adp} and Min\-kow\-ski \cite{hm1907k}. 
As usual, the natural blend of scientific curiosity, necessities arising from new engineering applications, and improvements in experimental techniques kicked off and drove the 
evolution that, in the course of the years until nowadays, brought interest and attention to spread to and explore other sectors of the vast physical phenomenology in question.
Impressive surveys, equipped with thorough collections of bibliographic references, are provided by Sagis \cite{ls2011rmp} and Somorjai \cite{gs2011pnas} with a view to terrestrial applications.

With the advent of the space age, aerothermodynamics (ATD) joined the collection of engineering applications for which the media-in-contact phenomenology assumes a role, and a rather crucial one in this case.
In ATD parlance, the term \textit{gas-surface interactions} (GSI) identifies the complex phenomenology of physical processes that occur inside the extremely thin transition layer between  hypersonic flows in thermo-chemical nonequilibrium and the walls of either a spacecraft during planetary (re)entry or a test probe in high-enthalpy wind tunnels. 
These physical processes produce macroscopically the \textit{superficial} exchange and transport of mass, momentum, energy and play a key role for the determination of several physical-variable distributions in both volume phases (fluid and solid) that turn out to be of utmost importance for heat-shield design. 
Among those physical variables, the wall heat flux is undoubtedly the most critical to aerothermodynamicists' concern. 

The theoretical understanding and physical as well as numerical modeling of the superficial processes is essential 
for establishment and exploitation of the correct boundary conditions necessary for the (numerical) solution of the field equations governing the volume phases.
Since the dawn of ATD, the GSI phenomenology has been approached in empirical manners, seemingly different in level of sophistication, that attempt to characterise the workings of the superficial processes by \textit{accommodation coefficients} (ACs), whose conceptual introduction is claimed to Knudsen \cite{mk1911adp} by Goodman and Wachman \cite{fg1976} or to Langmuir \cite{il1922tfs} by Billing \cite{gb2000} and Zangwill \cite{az1988}, and by-product parameters derived therefrom. 
Study, development and application of AC-structured models 
\cite{ja1989,pb2006jtht,mb1994aiaa,mb1996jsr,gb2015esa,sdb2010jtht,mf2015esa,ff2007jtht,nj2015csj,ak1999nasa,ak2000rtob,ak2009esa,ak2004iwpp,ak2002esa,ak1998esa,vk2005fd,vk1996fd,fn1996jtht,rtoenavt142,jp2009jtht,cp1990,is2015jsr,cs1985jsr,cs1992ah,jt2008ctr,av2007aiaa}
have proliferated in the course the 60 years elapsed from the appearance of the pioneering papers by 
Lees \cite{ll1956jjp}, Fay and Riddell \cite{jf1958jas}, Probstein \cite{rp1956jjp} and Goulard \cite{rg1958jjp}; the process is still ongoing nowadays. 
We refer readers wishing to acquire deeper familiarity with the subject matter to the thorough surveys of methods and literature provided by Kovalev and Kolesnikov \cite{vk2005fd} and Viviani \cite{av2008tn1}.
Although there are attempts \cite{mc1998aiaa,mc1999jtht,ic2007aiaa,mr2015jtht,mr2009jpc,cz2012jpc} to determine ACs from quantum-me\-chan\-i\-cal, molecular-dy\-nam\-ics {\color{black} (MD)} and Monte-Carlo calculations, efforts and energies (financial form included) have been invested prevalently in experimental investigations \cite{mb1997ass,mb2003cp,mb2010ass,lb2005ast,be2015esa,gh2005jsr,nj2015csj,ak2000rtoa,ak2000jsr,ak1999aiaa,ak1998esa,bm2015esa,bm2015asr,fp2012mcp,fp2011aiaa,fp2014ass,sp2005jtht,ms2006iac}.  
As a matter of consequence, empirical GSI models incorporate implicitly experimental information, confidently accurate for sure from the standpoint of ex\-per\-i\-men\-tal-testing art but of unexplored and therefore questionable physical significance, to provide \textit{realistic} (?) numbers for ACs and related variables.
Indeed, and experimental imprimatur notwithstanding, there is still a broad variety of opinions among aerothermodynamicists regarding the correct and, a\-bove all, \textit{convincing} answer to a several-decade old but still lingering question: what \textit{independent} physical variables do the ACs depend on?
Collectable answers from workers in the field are manifold and, regrettably, contradictory.
After almost three de\-cades (in Europe) of AC practice and operations, it seems (to us) fair to admit that the theoretical/nu\-mer\-i\-cal state of the AC art is today still rather far away from deserving satisfactory engineering confidence.
The main reason for the virtual distance to the latter target lies in the true nature of the AC-concept: it is a heuristic, empirical construct lacking support from an underlying rigorous and self-contained physical theory, as explicitly declared also by Kovalev and Kolesnikov \cite{vk2005fd}.
Broadly speaking, ACs and derived by-products are nothing more than empirical parameters whose usage is a kind of numerical subterfuge to avoid paying attention to the complex details of the physical processes going on inside the transition layer.
AC methodologies are habitually trusted with owning predictive power and capabilities to shortcut, or even elude, superficial-process complexities, presumptively conferred on them
without perhaps realising the great risk lurking behind that conferral: the in-vain tentative to describe in an \textit{algebraic} manner a physical phenomenology that is governed by differential equations whose boundary conditions' variety pulverises any however perceptibly small allusion to the existence of algebraic functions for the ACs.  
In our opinion, the AC-modeling pathway is conceptually unsatisfactory, inherently difficult and rather uncontrollable as far as accuracy is concerned, mainly because experimental testing with high-tem\-per\-ature flows in thermochemical non-equilibrium is costly and not as thoroughly feasible as one would wish for the purpose of explaining through the AC-modeling pathway all the open question marks of the GSI phenomenology; as a result, prediction reliability and engineering effectiveness of AC-based empirical GSI models could be se\-vere\-ly compromised. 

After slightly more or less than \textit{one century} since Knudsen \cite{mk1911adp} and/or Langmuir \cite{il1922tfs} proposed the AC concept and with ATD applications in mind,
it appears (to us) unavoidable to confront an obvious, and maybe hopeless, question: will any major breakthrough be ever achieved by obstinate persistence on the AC-modeling pathway?
The majority of aerothermodynamicists has resigned to persistence and endures case-by-case engineering challenges, and perhaps justifiably so because \textit{at the moment} there is no other alternative compatible with the rigid time frames imposed by heat-shield design. 
However, a few of them, lucky enough to be unconstrained by design necessities, believe time is mature to, at least, explore other scientific avenues.

\section{Theoretical considerations}

The correct mathematical description of interfacial physics presupposes a clear understanding of what happens at microscopic level \cite{ta2002ap,gb2000,fg1976,ag2009,rm2013,gs1994,az1988}  in the transition layer.
As simple and convenient entry point into the discourse, let us consider two stationary solids in contact as sketched in \Rfi{mvmsc}.
\begin{figure}
  \fbox{\includegraphics[bb = 160 120 515 760 , clip , width=.975\columnwidth]{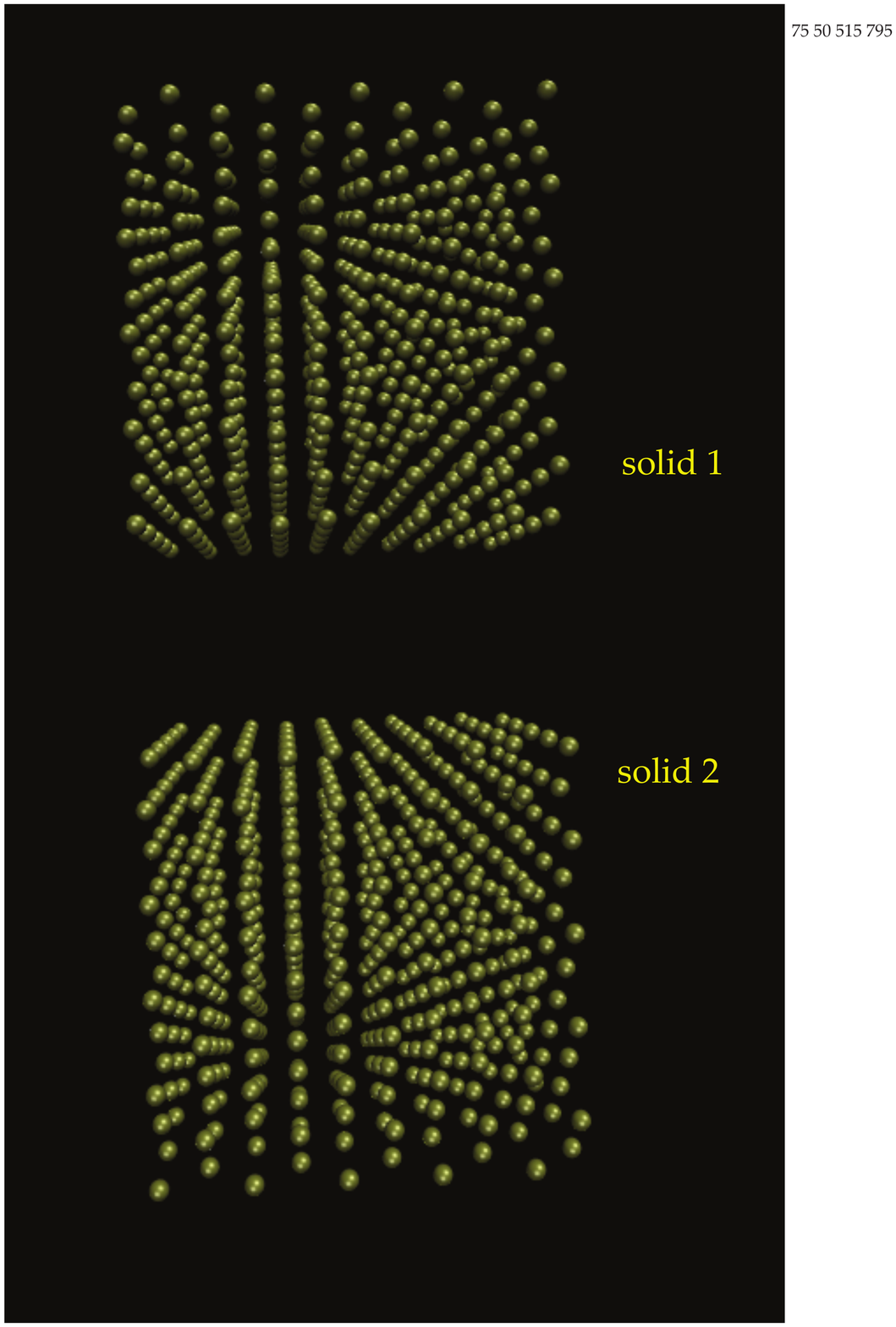}}
  \caption{Microscopic view of two blocks of platinum in contact at a distance of 4 $\mathring{\mbox{a}}$ngstroms. Courtesy of Prof S V Nedea, dpt of Mechanical Engineering, Technical University of Eindhoven}
  \label{mvmsc}
\end{figure}
Obviously, the \textit{contact} concept is macroscopic in nature because the atoms in the terminal rows of the solids do not really \textit{touch}; they stand off at a minute distance established by the interplay between the forces due to the repulsive potential that comes to exist in the region where the terminal rows are facing eachother and other external forces imposed on the solids.
Exchanges of momentum and energy between the solids occur through the mediating action of the potential, unconditionally modelled as action-at-distance-like (for simplicity, let us imagine also radiation negligible and put it aside).
It appears evident that, under this circumstance, momentum and energy leaving one solid per unit area in the unit time can only go into the other solid. 
Thus, the ther\-mo\-dy\-nam\-ic-energy exchange in heat-transfer problems is consistently subjected macroscopically (\Rfi{Mvmsc})
\begin{figure}
  \fbox{\includegraphics[bb = 30 120 585 675 , clip , width=.975\columnwidth]{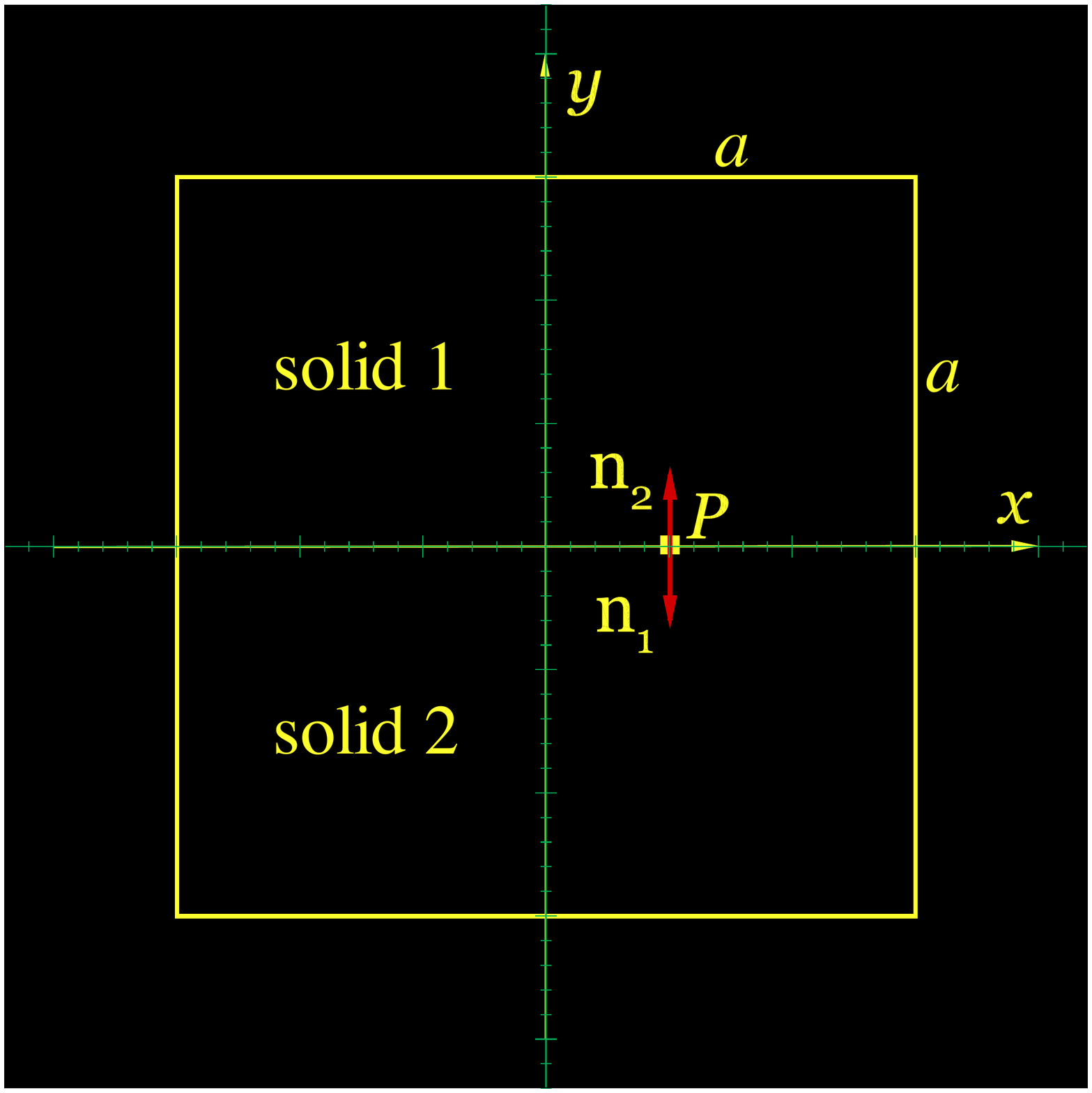}}  
  \caption{Macroscopic representation of the platinum contact sketched in \Rfi{mvmsc}}
  \label{Mvmsc}
\end{figure}
to heat-flux continuity
\begin{equation}
  \nuv{1} \cdot \df{U,1}(P,t) + \nuv{2}\cdot \df{U,2}(P,t)= 0
  \label{hfc}
\end{equation}
in any instant of time $t$.
\REqb{hfc} is a straightforward consequence of the principle of total-energy conservation, although only the thermodynamic form $U$ is at play in this case. 
However, heat-flux continuity alone is not sufficient to describe the heat-transfer dynamics. 
Another physical condition is needed and temperature continuity 
\begin{equation}
  T_{\tsub{1}}(P,t) - T_{\tsub{2}}(P,t) = 0
  \label{Tc}
\end{equation}
(or even its more sophisticated contact-resistance version) turns out to be customarily enforced.
\REqb{Tc} is a typical example of intuitively imposed \textit{ad-hoc} condition.
Indeed, although a widely exploited textbook standard \cite{yc2015,rh2009}, \REq{Tc} becomes conceptually a bit problematic when the underlying justifying physical principle is sought for.
In fact, there is none. 
Sometimes vague claims appealing to (typically unreferenced) experimental evidence are invoked to support the unconditional enforcement of \REq{Tc} but they are easily contrasted and dismissed by documented experimental evidence showing the opposite as, for example, the results reported by Fang and Ward \cite{gf1999pre} for a liquid-vapour interface. 
We will come back to this interesting aspect in the sequel (\Rse{httc-std}).

If the external forces are sufficiently intense, the microscopic situation turns into the one sketched in \Rfi{mvmsci}: the terminal rows of atoms compenetrate and form a thin layer of microscopic thickness that cannot be considered either solid 1 or solid 2. 
\begin{figure}
\framebox{\includegraphics[bb = 100 40 365 500 , clip , width=.975\columnwidth]{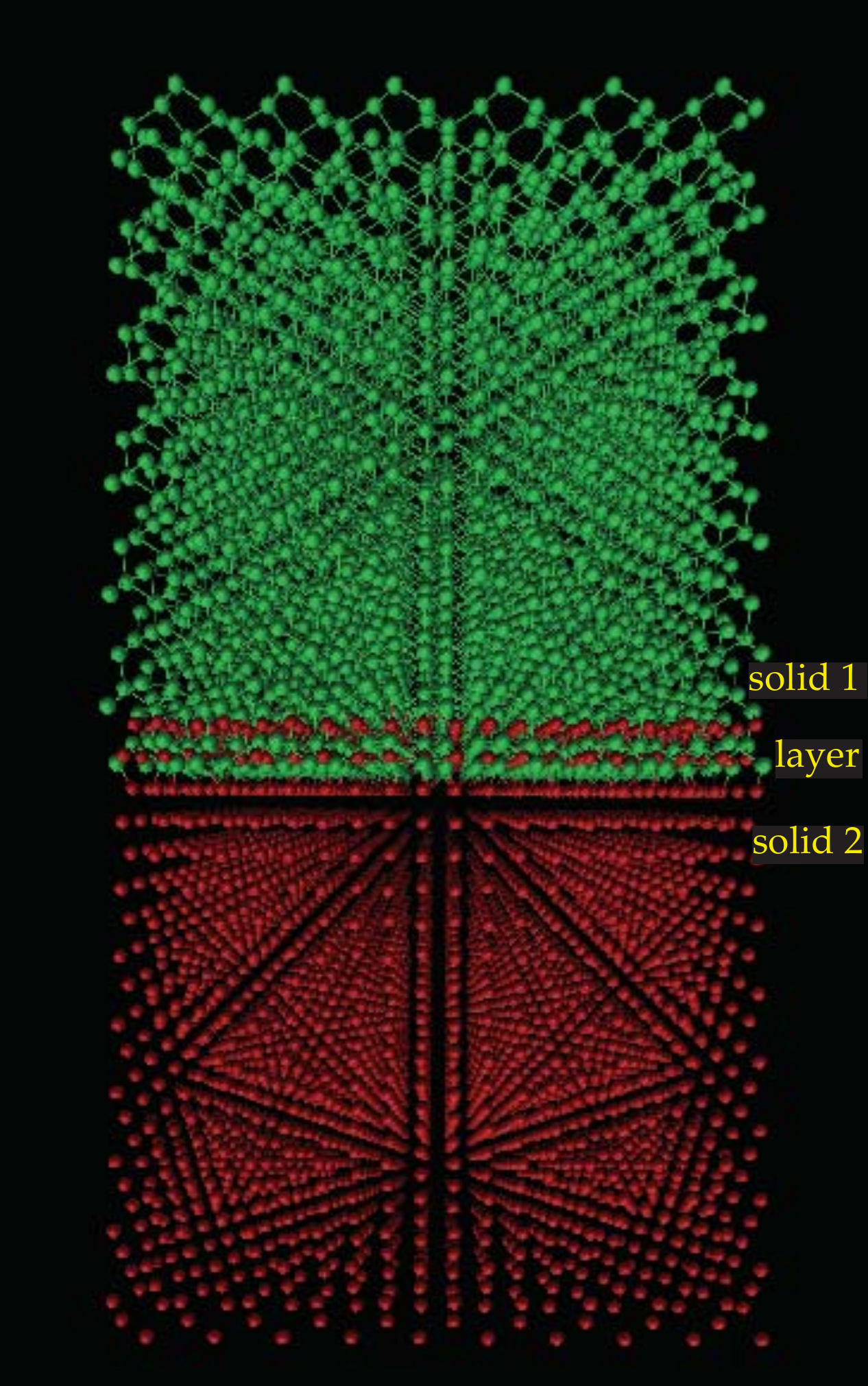}}
\caption{Microscopic view of a block of silicon (solid 1) and a block of platinum (solid 2) separated by a thin mixed layer. Courtesy of Prof S V Nedea, dpt of Mechanical Engineering, Technical University of Eindhoven}
\label{mvmsci}
\end{figure}
\begin{figure}  
  \fbox{\includegraphics[bb = 30 120 585 675 , clip , width=.975\columnwidth]{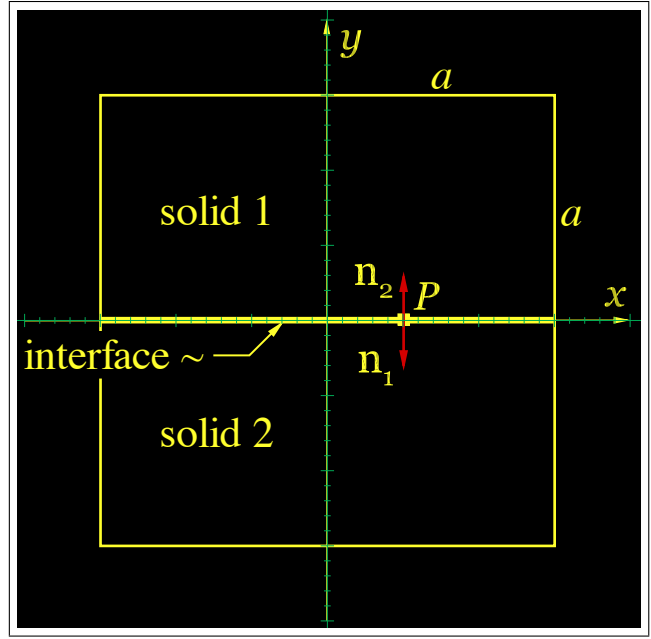}}  
  \caption{Macroscopic representation of the silicon-platinum contact sketched in \Rfi{mvmsci}; planar interface.}
  \label{Mvmsci}
\end{figure}
\begin{figure}
  \fbox{\includegraphics[bb = 9 99 603 694 , clip , width=.975\columnwidth]{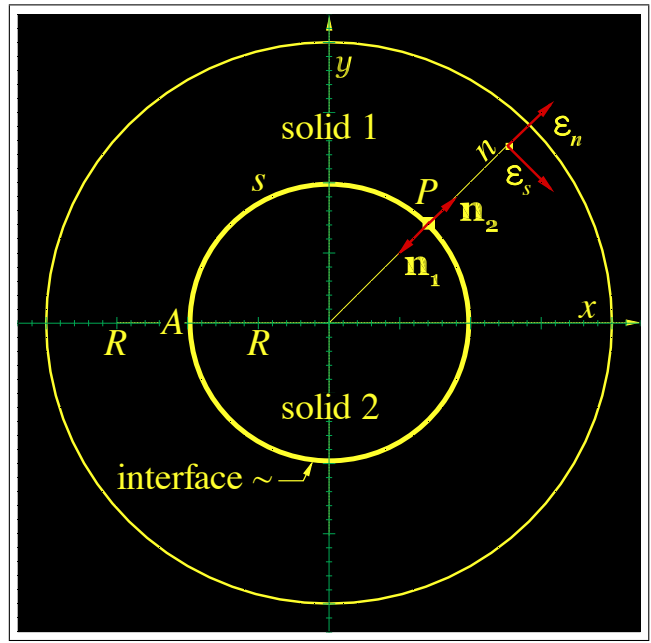}}  
  \caption{Macroscopic representation of two-solid cylindrical contact; curved interface.}
  \label{Mvmsci.c}
\end{figure}
Physical properties leaving one solid per unit area in the unit time do not go directly into the other solid but have to cross the thin layer first, and it is clear that, therein, they can be locally accumulated, produced, and, rather interestingly, \textit{tangentially} transported, by diffusion in this case.
In other words, the thin layer that forms between the media in contact turns out to be a third medium with its own physical properties (mass, momentum, energy in its various forms and total, etc) that follow their own evolutional dynamics.
With this microscopic picture in mind, the foundational idea on which a phenomenological theory can be built is to model the thin layer macroscopically as a \textit{two}-dimensional continuum, a \textit{surface phase} or \textit{interface} in between the volume phases, characterised by physical properties per unit area.
Since Gibbs introduced the embryonic form of this idea in his seminal memoirs \cite{jg1876tca,jg1878tca,jg1878ajs}, also included in his collected works \cite{jg1906v1,jg1928v1}, within a thermodynamics context limited to energy and entropy, it has been pursued, explored and made evolve during the course of the years again in thermodynamics \cite{fb1960sl,rd1977jcis,rg2001,eg1940tfs,om2011,ln1978aiaa,so1960sl,as1968} as well as in kinetic theory {\color{black} (KT)} \cite{vb1988sp,vb1990sp,sk1999ss,sk1997jcp}, linear irreversible thermodynamics (LIT) \cite{aa1987p,db1986acp,db1976p,sk2008,rg2001,jk1977p,ln1978aiaa,ho2009pre,ls2011rmp,lw1967zfn}, and phe\-nom\-e\-no\-log\-i\-cal theory 
\cite{fdi1993am,rg2001,vg1983fd,mg1998pma,aj2014cmame,aj2013amr,vl1969arfm,ln1978aa,ln1979aa,ap1979m,rp1971jp,rp1976,ls2011rmp,ls1960ces,js1964ces,js1967iecf,js2007};
fa\-mil\-iar\-isation with the mentioned literature, and that contained there\-in, is indispensable for those who wish to endeavour to work in this scientific domain.

The evolution of the superficial physical properties is governed by superficial balance equations that play a bridging role between the field equations governing the evolution of the physical properties belonging to the volume phases.
The three sets of equations are obviously coupled and require a simultaneous (numerical) solution.
We recommend readers interested in specific details and rigorous mathematical derivations to consult the mentioned literature, in particular 
Bedeaux et alii \cite{db1986acp,db1976p}, Gatignol and Prud'homme \cite{rg2001}, Gogosov et alii \cite{vg1983fd}, Napolitano  \cite{ln1978aiaa,ln1978aa,ln1979aa}, 
Slattery \cite{js1964ces,js1967iecf}, and Slattery et alii \cite{js2007};  
our notation and terminology will follow very closely those adopted by Napolitano.
The mathematical structure of the superficial balance equations is conceptually very similar to the familiar one of the balance equations applicable in the volume phases; 
the formal one corresponding to the generic scalar or vectorial extensive physical variable $G$ reads
\begin{equation}
   \pder{\gta}{t} + \Dives[\Vt\gta-\nuv{}\times(\nuv{}\times\dft{G})] =
   \nuv{1}\cdot\tf{G,1} + \nuv{2}\cdot\tf{G,2} + \gtpa
   \label{sbe}
\end{equation}
In \REq{sbe},
\begin{tabbing}
  xx \= xxxxx \= xxxxxxxxxx \kill
  \> $\gta$       \> superficial density \\
  \> $\Grads$     \> surface-gradient operator \\
  \> $\Vt$        \> superficial mass velocity \\
  \> $\nuv{k}$    \> surface-normal unit vector for volume phase\\[-.25\baselineskip]
  \>                 \> $k$ (=1,2); chosen outwards by convention\\
  \> $\nuv{}$     \> surface-normal unit vector (either of the two)\\
  \> $\dft{G}$    \> superficial diffusive flux \\
  \> $\tf{G,k}$ \> total flux in volume phase $k$ \\
  \> $\gtpa$      \> superficial production 
\end{tabbing}
\REqb{sbe} applies locally in each geometrical point of the interface.
The term
\begin{equation}
  \nuv{1}\cdot\tf{G,1} + \nuv{2}\cdot\tf{G,2}
  \label{tfj}
\end{equation}
on the right-hand side is the total-flux jump across the interface and represents the main, although not the only one, channel of physical-property exchange between the interface and the volume phases. 
The total flux in volume phase $k$ is obviously separable in convective and diffusive contributions
\begin{equation}
  \tf{G,k} =\rho_{\tsub{k}} ( \V_{\tsub{k}} - \W ) g_{\tsub{k}} + \df{G,k}
  \label{tf3d}
\end{equation}
with
\begin{tabbing}
  xx \= xxxxx \= xxxxxxxxxx \kill
  \> $\rho_{\tsub{k}}$ \> mass density \\
  \> $\V_{\tsub{k}}$   \> mass velocity \\
  \> $\W$                \> interface \textit{geometrical} velocity \\
  \> $g_{\tsub{k}}$     \> volume density\\
  \> $\df{G,k}$         \> diffusive flux
\end{tabbing}
\textcolor{black}{Of course, the superficial balance equations [\REq{sbe}] are in open form and require to be complemented with phenomenological relations for superficial diffusive fluxes and productions, exactly as it happens for the volume-phase balance equations. 
The phenomenological relations are expected to be provided by joint efforts and combined contributions from {\color{black} MD}, thermodynamics, LIT, and KT \textit{of the interface}, desirably supported by experimental verification, if any be possible. 
In this regard, there is a noticeable volume of knowledge already available in the literature mentioned in the two paragraphs above \REq{sbe}.}
The concept of \textit{two-dimensional fluid dynamics} may sound exotic to some of us but this intuitive (perhaps even emotional sometimes) reaction is just a consequence of mental habit acquired in years, educational and professional, of dealing and practicing with applications in three-dimensional space behaving marvelously well according to the Euclidean prescript.
The habit buildup starts early in our formation days: in general, fluid-dynamics textbooks default, without deliberate and explicit recognition, to such a spatial circumstance and unravel physical concepts in mathematical formalisms framed on systems of orthogonal coordinates among which the Cartesian ones are particularly more privileged than the curvilinear ones.
There is a laudable exception though: Sedov \cite{ls1965,ls1966,ls1971,ls1975v1} teaches continuum mechanics in Riemann space and uses co/contra-variant non-orthogonal curvilinear coordinates throughout (a really mind-broadening learning experience!).
An additional setback is the absence of two-di\-men\-sional fluid dynamics  in universities' syllabuses and the systematic lack of consideration for it in textbooks.
But also in this regard there is a commendable exception: Aris \cite{ra1989}, after preliminarly introducing ``the geometry of surfaces in space'' (chapter 9), provides an interesting treatment of ``the equations of surface flow'' (chapter 10) admittedly inspired and based on Scriven's work \cite{ls1960ces}. 
In the paragraph introducing the latter chapter, he emphasises straight off two among the most important and crucial characteristics of interfaces.
The first:
\begin{quote}
 \ldots Cartesian tensors really suffice for three-di\-men\-sion\-al flows; for the space of everyday life, being Euclidean, always admits of a Cartesian frame of reference.
 However, the surface is a two-dimensional non-Euclidean space and from the outset demands a full tensorial treatment.
 \end{quote} 
This characteristic follows by necessity from the fact that the 
\begin{quote}
 \ldots surface is a two-dimensional space that can move within a space of higher dimensions, namely, the three-dimensional space surrounding it.
 \end{quote} 
Thus, position and shape of the interface are, in general, time dependent and not known \textit{a priori}; in other words, the geometrical equation of the interface 
\begin{equation}
  \Rs = \Rs(\gc{1},\gc{2},t) 
\end{equation}
is an unknown of the problem and, consequently, the Gaussian curvilinear coordinates $(\gc{1},\gc{2})$ cannot be assumed unconditionally orthogonal from the outset.
The necessity of a full tensorial treatment is so essential that Napolitano \cite{ln1977ams} dedicated a full paper and Gatignol and Prud'homme \cite{rg2001} a thorough appendix to the subject matter although, peculiarly enough, these authors target only orthogonal curvilinear coordinates. 
Incidentally, the motion of the interface affects explicitly the fluxes [\REq{tf3d}] through the presence of the geometrical velocity defined as
\begin{equation}
  \W = \Pder{\Rs}{t}{\gc{1},\gc{2}} 
  \label{gv}
\end{equation}
The second characteristic
\begin{quote}
 \ldots [the interface] may be the region of contact of two bulk fluids. This is again a new feature, for a bulk fluid can never be the interface of two four-dimensional fluids.
\end{quote} 
follows from the presence of the total-flux jump [\REq{tfj}], a property belonging to the volume phases, in the superficial balance equations [\REq{sbe}].
It is the element that appoints the superficial balance equations to the role of boundary conditions, an aspect specifically discussed in Sec. 10.51 (``Surface equations as boundary conditions at an interface''), also prototyped by Waldmann \cite{lw1967zfn} and repeatedly stressed by Bedeaux et alii \cite{db1986acp,db1976p} and Napolitano \cite{ln1978aiaa,ln1978aa,ln1979aa}.
In this regard, we should not miss an essential point already mentioned in \Rse{intro}: the physical-property exchanges through the interface constitute a physical phe\-nom\-e\-nol\-o\-gy governed by differential equations [\REq{sbe}] and any attempt to describe it algebraically, as done along the AC-modeling pathway, is hopelessly ill-fated.

In conclusion, it should appear evident from the previous considerations that a phenomenological theory of interfaces targeting non-space applications has been maturing since long time.
The novelty of our research program is the tentative extension of such a phenomenological theory to build a branch, a macroscopic theory of gas-surface interactions (MTGSI), targeting the complex phenomenology featured by hypersonic reentry flows as alternative to the AC-modeling pathway.
\begin{figure}[h]
	\framebox{\includegraphics[bb = 1 0 795 420 , clip , width=.975\columnwidth]{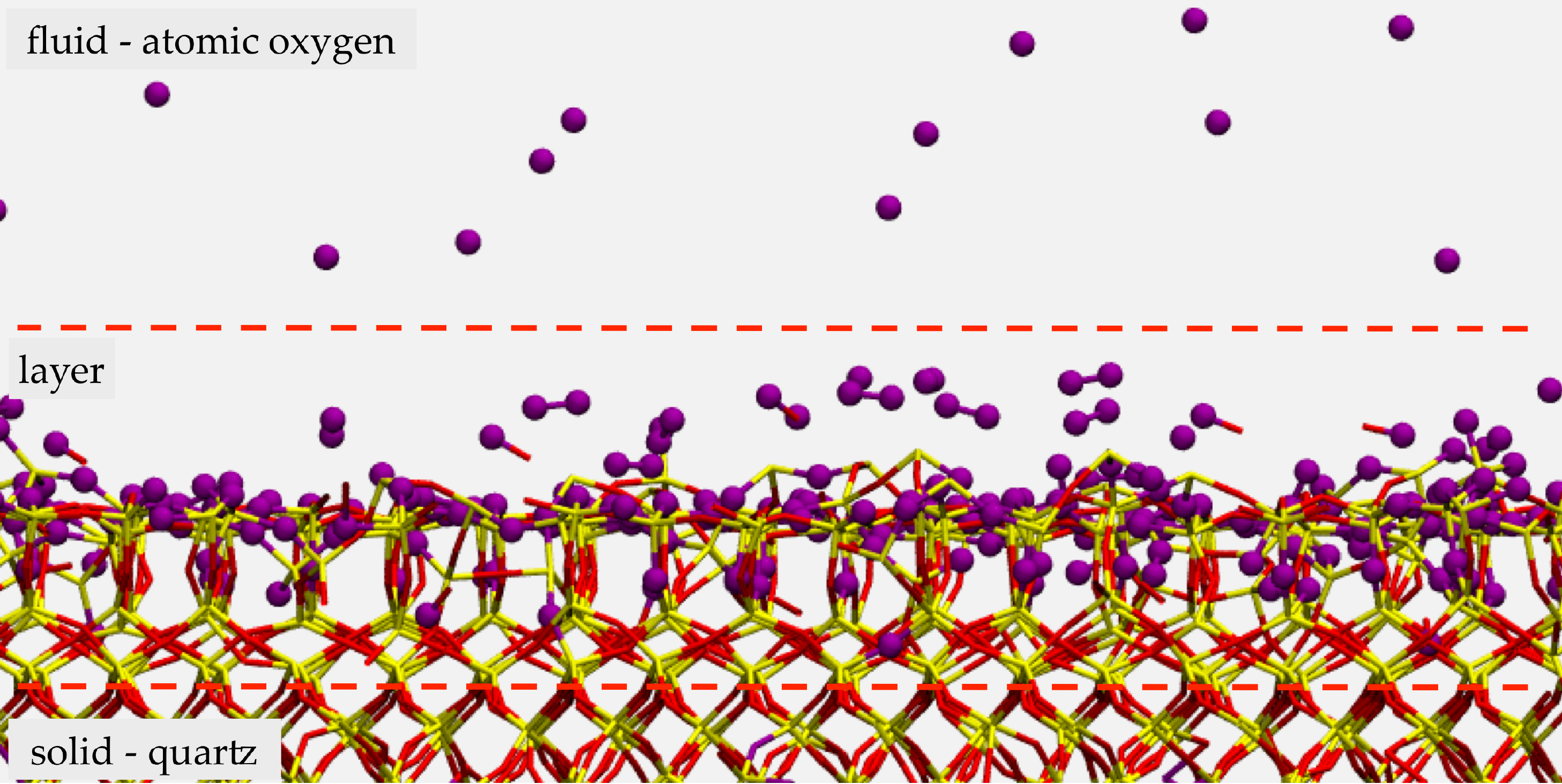}}
	\caption{Microscopic view of DSMC-reconstructed quartz surface exposed to atomic oxygen. Courtesy of Prof T Schwartzentruber, dpt of Aerospace Engineering and Mechanics, University of Minneapolis}
	\label{mvmsci-fd}
\end{figure}
\begin{figure}
	\fbox{\includegraphics[bb = 10 100 600 694 , clip , width=.975\columnwidth]{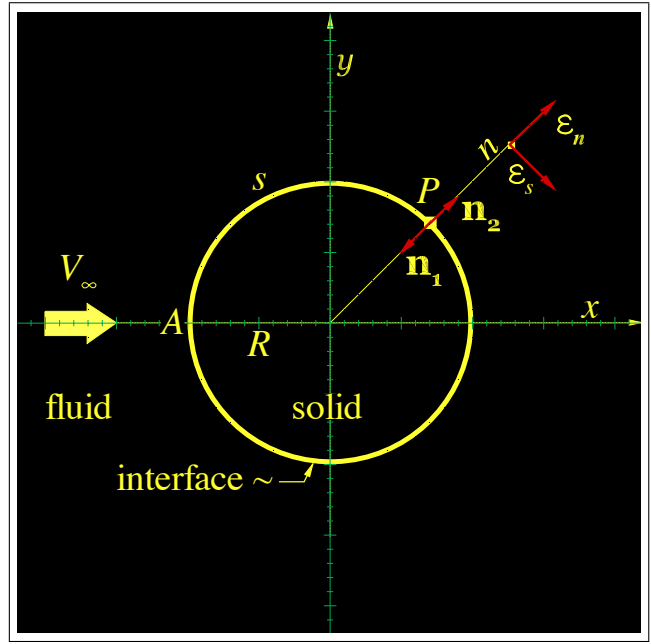}}  
	\caption{Future study: two-dimensional flow past a cylinder with (and without) interface.}
	\label{Mvmsci.c.fd}
\end{figure}
Clearly, the idea of interface formation, and its macroscopic modelling as a two-dimensional continuum, can be smoothly exported without any particular conceptual difficulty to the case of fluid-solid contact;  examples of microscopic view and a possible macroscopic configuration are sketched in \Rfid{mvmsci-fd}{Mvmsci.c.fd}.
Nevertheless, the achievement of the preparedness to deal theoretically, mathematically and numerically with the global set of coupled governing equations for fluid, solid and interface appears a rather ambitious feat at this stage of our research programme.
\setlength{\marginparwidth}{.5\marginparwidth}
{\color{black} This sensation becomes even more acute in view of confronting the very complex physical phenomenology embraced by aerothermodynamics. 
The phe\-nomenological-relation know-how ac\-quired for non-space problems [see bibliographic references mentioned before \REq{sbe}] will certainly require further advancement to shed light on and to bring within reach the complexities of not only familiar, although still ill-understood, processes such as superficial chemical kinetics and ablation but also of others playing a role in \REq{sbe} and ignored right now because they are invisible along the AC- and ablation-modeling pathways.
As explicit example, and a rather significant one by the way, let us scrutinise the balance 
\definecolor{gray}{gray}{0.7}
\begin{equation}
  \begin{split}
     \color{black}
     \pder{\rta_{\tsub{i}}}{t} + \Dives[\rta_{\tsub{i}}\Vt & \color{black}-\nuv{}\times(\nuv{}\times\dfmt{i})]   = \\
     & \quad \quad \;\nuv{1}\cdot[\rho_{\tsub{i,1}} ( \V_{\tsub{1}} - \W )  + \underline{\dfm{i,1}}] \\[.5\baselineskip]
     & \quad \color{black}+ \nuv{2}\cdot[\rho_{\tsub{i,2}} ( \V_{\tsub{2}} - \W ) + \dfm{i,2}] \\[.5\baselineskip]
     & \quad + \underline{\textstyle\sum_{\tsub{r}} \crrt{r} \nus_{\tsub{ri}} M_{\tsub{i}}}
  \end{split}
\label{sbe.cc}
\end{equation}
of the superficial mass of a chemical component. 
In \REq{sbe.cc}, the subscripts $i$ and $r$ point respectively to chemical component and chemical reaction, and
\begin{tabbing}
	xx \= xxxxx \= xxxxxxxxxx \kill
	\> $\rta_{\tsub{i}}$     \> superficial partial density \\
	\> $\dfmt{i}$             \> superficial-mass diffusive flux \\
	\> $\rho_{\tsub{i,k}}$  \> volume-phase partial density \\ \>  \> ($k=1$ is gas, $k=2$ is solid) \\
	\> $\dfm{i,k}$           \> volume-phase mass diffusive flux \\[.25\baselineskip]
    \> $\crrt{r}$              \> superficial chemical-reaction rate \\ 
    \> $\nus_{\tsub{ri}}$    \> stoichiometric coefficient \\ 
    \> $M_{\tsub{i}}$        \> molecular mass \\ 
\end{tabbing}
AC-structured models shut convective fluxes and solid-side diffusive flux on the right-hand side of \REq{sbe.cc} out of the picture and contemplate exclusively the interplay between the underlined terms that describe, respectively, gas-side mass diffusion, systematically assumed governed by the poorly-performing Fick law, and superficial chemical kinetics; in turn, ablation models 
\cite{db2007,db2012jpp,db2009jsr,db2011jsr,db2011aiaa,jk1993aiaa,jk1994aiaa,rk1968nasa,fm1994jtht,at2013}
allow the convective fluxes to operate.
In so doing, those models disregard mass diffusion in solids, which is not an unknown phenomenology  \cite{ea1980am,rb1951,af1855adp,hm2007,ps2016}, and, more importantly, miss the full extent of role played and consequences produced by the terms appearing on the left-hand side of \REq{sbe.cc}, particularly those that bring into account convective ($\rta_{\tsub{i}}\Vt$) and diffusive ($\dfmt{i}$) transport \textit{tangentially within} the interface.
With specific regard to AC-structured models, there is no amount of squeezing \textit{algebraic} coefficients into the expressions of the superficial chemical-reaction rates ($\crrt{r}$) that will substitute for the left-hand-side terms and will establish physical equivalence with and prediction capability of the full \textit{differential} equation.

The previous example provides sufficient hints of the broader scope of the MTGSI-mod\-eling pathway and also of the conceptual hurdles disseminated along it;  we need not to delve further into the meanders of the superficial balance equations of other physical properties such as total mass, momentum, etc., to convince ourselves that the list of conceptual corners awaiting illumination and exploration is obviously rather long.}
We wish, of course, to undertake {\color{black} the} self-learning {\color{black} exploration} walk-through in steps of increasing difficulty.
{\color{black} As} start-up of our research program, {\color{black} therefore,} we concentrate on the exploratory study of the heat-transfer test case mentioned in the beginning of this section in planar and cylindrical configurations with and without interface.
It should be looked at as a numerical demonstrator, simple enough to allow quick numerical computations but already sufficiently meaningful to bring forth important aspects relevant to thermal protection arising from the presence of an interface; to some extent, it also brings to completion the work initiated by Schmidtmann \cite{bs2013vki}.
As by-product, we show that, under circumstances of interface absence as sketched in \Rfi{mvmsc}, the popular temperature-continuity boundary condition [\REq{Tc}] turns out to be a particular case whose customarily assumed unconditional applicability is not justified by any physical principle of conservation.

\section{Heat-transfer test case without interface} \label{httc-std}

\subsection{Theoretical considerations}
In the case of absent interface ($\gta\rightarrow 0, \dft{G}\rightarrow 0$), the surface balance equation [\REq{sbe}] for a conservative ($\gtpa=0$) variable $G$ reduces to the familiar total-flux jump condition
\begin{equation}
   \nuv{1}\cdot\tf{G,1} + \nuv{2}\cdot\tf{G,2} = 0
   \label{jc}
\end{equation}
that, for solids as those of \Rfi{mvmsc} for example ($\V_{\tsub{k}}=\W=0$), simplifies even further to involve only the diffusive fluxes
\begin{equation}
   \nuv{1}\cdot\df{G,1} + \nuv{2}\cdot\df{G,2} = 0
   \label{jc-df}
\end{equation}
The specialisation of \REq{jc-df} to total mass ($G=m$) is identically satisfied because total mass cannot diffuse ($\df{m,k}=0$) by definition.
The specialisation of \REq{jc-df} to thermodynamic energy ($G=U$) reproduces the heat-flux continuity [\REq{hfc}] that we have already considered in \Rse{intro}.
The specialisation of \REq{jc-df} to momentum ($G=\bv{Q}$) leads to tension (force per unit area) continuity
\begin{equation}
   \nuv{1}\cdot\st{1} + \nuv{2}\cdot\st{2} = 0
   \label{mfc}
\end{equation}
In \REq{mfc}, $\st{\tsub{k}} = - \df{\bv{Q},k}$ is the stress tensor in solid $k$.
If we take into account that stresses in solids depend on local temperature \cite{hc1963,rh2009} then we understand right away that \REq{mfc} contains and provides the other condition that, together with heat-flux continuity [\REq{hfc}], governs the establishment of the temperature difference $T_{\tsub{1}}(P,t) - T_{\tsub{2}}(P,t)$ (\Rfi{Mvmsc}); 
thus, the \textit{physical} tension-continuity condition gently relegates the \textit{intuitive} tem\-per\-a\-ture-continuity condition [\REq{Tc}] to the role of particular case.

In order to make these theoretical considerations explicit with simple test cases, and for the purpose of easiness in numerical calculations, we assume the solids to be isotropic, to follow Fourier law 
\begin{equation}
  \df{U,k} = - \tc{k} \Grad T_{\tsub{k}}
  \label{hf-fl}
\end{equation}
and to feature a tensional behaviour described by Hooke law with linearised thermal-stress contribution \cite{rh2009}
\begin{equation}
  \st{k} = \st{e,k} - \beta_{\tsub{k}} ( T_{\tsub{k}} - T_{\tsub{0,k}} ) \ut \simeq - \beta_{\tsub{k}} ( T_{\tsub{k}} - T_{\tsub{0,k}} ) \ut 
  \label{hl-lts}
\end{equation}
In \REqd{hf-fl}{hl-lts}
\begin{tabbing}
  xx \= xxxxx \= xxxxxxxxxx \kill
  \> $\tc{k}$       \> thermal conductivity in solid $k$ (=1,2) \\
  \> $\Grad$       \> volume-gradient operator \\
  \> $\st{e,k}$     \> stress-tensor elastic part \\
  \> $\beta_{\tsub{k}}$   \> thermal-stress coefficient \cite{hc1963} \\
  \> $T_{\tsub{0,k}}$     \> reference temperature \\
  \> $\ut$           \> unit tensor 
\end{tabbing}
The thermal-stress coefficients can be expressed as
\begin{equation}
  \beta_{\tsub{k}} = \frac{ E_{\tsub{k}} }{ 1 - 2\nu_{\tsub{k}} }  \alpha_{\tsub{k}}
  \label{tsc}
\end{equation}
In \REq{tsc}
\begin{tabbing}
  xx \= xxxxx \= xxxxxxxxxx \kill
  \> $E_{\tsub{k}}$         \> Young modulus \\
  \> $\nu_{\tsub{k}}$       \> Poisson ratio \\
  \> $\alpha_{\tsub{k}}$   \>  linear thermal-expansion coefficient \cite{rh2009} or \\
  \>                    \>  thermal-strain coefficient \cite{hc1963}
\end{tabbing}
Thermodynamic and transport properties are scalar as a consequence of the isotropy of the solids and are assumed constant for simplicity.
In \REq{hl-lts}, we consider negligible the stress-tensor elastic part with respect to the thermal-stress contribution for the simple reason of avoiding to consider deformations in the solids.  
The inclusion of stress-tensor elastic part and deformation field is, of course, doable but introduces the numerical complicacy of solving the deformation field together with the temperature field, an unnecessary sophistication in the context of the present discourse and that, by itself, does not add or remove any physical significance to the numerical results described in the following sections.

On the basis of the previous set of assumptions, \REq{hfc} particularises to
\begin{equation}
   \tc{1} \nuv{1} \cdot \Grad T_{\tsub{1}}|_{\tsub{P,t}} + \tc{2} \nuv{2}\cdot \Grad T_{\tsub{2}}|_{\tsub{P,t}} = 0
  \label{hfc-fl}
\end{equation}
and \REq{mfc} reduces to a vector equation only in the normal direction ($\nuv{k}\cdot\ut = \nuv{k}$)
\begin{equation}
   \beta_{\tsub{1}} [ T_{\tsub{1}}(P,t) - T_{\tsub{0,1}} ] \nuv{1} + \beta_{\tsub{2}} [ T_{\tsub{2}}(P,t) - T_{\tsub{0,2}} ] \nuv{2}\simeq 0
   \label{mfc-hl-lts-v}
\end{equation}
from which we obtain the scalar condition ($\nuv{2}= - \nuv{1}$)
\begin{equation}
   \beta_{\tsub{1}} [ T_{\tsub{1}}(P,t) - T_{\tsub{0,1}} ] - \beta_{\tsub{2}} [ T_{\tsub{2}}(P,t) - T_{\tsub{0,2}} ] \simeq 0
   \label{mfc-hl-lts-s}
\end{equation}
enforcing tension continuity in the normal direction.
Temperature continuity is recovered in the particular case when $\beta_{\tsub{1}}=\beta_{\tsub{2}}$ and $T_{\tsub{0,1}}=T_{\tsub{0,2}}$.

\subsection{Application to planar configuration} \label{std-pg}
With reference to \Rfi{Mvmsc}, we assume 
\begin{itemize}
     \item adiabatic vertical walls
	       \begin{equation}  \Pder{T_{\tsub{k}}}{x}{x=\pm a} = \Pder{T_{\tsub{k}}}{z}{z=\pm a} =0 \label{avw} \end{equation}
     \item uniform initial temperature
            \begin{equation} T_{\tsub{k}}(x,y,z,0) = T^{\tsup{i}} \label{ic} \end{equation}
     \item uniform temperature at the top of solid 1 increasing linearly in time, within a finite interval $\ttd$, from the initial value $T^{\tsup{i}}$ to a final value $T^{\tsup{f}}$ 
             \begin{equation}
                T_{\tsub{1}}(x,+a,z,t) = \left\{ 
                                              \begin{array}{l@{\hspace{1.5em}}c} T^{\tsup{i}} & \hspace{1.75em} t < 0 \\ 
                                                                   \displaystyle         T^{\tsup{i}} + (T^{\tsup{f}}-T^{\tsup{i}})\frac{t}{\ttd} &  0\leq t \leq \ttd \\ 
                                                                                          T^{\tsup{f}} & \hspace{-1.75em} \ttd < t \end{array}  \right.
                                                                   \label{topT}
              \end{equation}
      \item uniform temperature at the bottom of solid 2 kept at the initial value
             \begin{equation}  T_{\tsub{2}}(x,-a,z,t) = T^{\tsup{i}} \label{bottomT} \end{equation}
\end{itemize}
This set of boundary conditions [\REqs{avw}{bottomT}] determines a one-dimensional unsteady heat transfer in the $y$ direction governed by the standard diffusion equation
\begin{equation}
  \rho_{\tsub{k}} c_{\tsub{k}} \pder{T_{\tsub{k}}}{t} = \tc{k}\psde{T_{\tsub{k}}}{y} 
  \label{htge}
\end{equation}
In \REq{htge}
\begin{tabbing}
  xx \= xxxxx \= xxxxxxxxxx \kill
  \> $\rho_{\tsub{k}}$   \> density of solid $k$ (=1,2) \\
  \> $c_{\tsub{k}}$       \> specific heat 
\end{tabbing}
The boundary conditions at the separation surface [\REqd{hfc-fl}{mfc-hl-lts-s}] become
\begin{equation}
   \tc{1} \Pder{T_{\tsub{1}}}{y}{y=\zerop} - \tc{2} \Pder{T_{\tsub{2}}}{y}{y=\zerom}= 0
  \label{hfc-fl-pg}
\end{equation}
\begin{equation}
   \beta_{\tsub{1}} [ T_{\tsub{1}}(\zerop,t) - T^{\tsup{i}} ] - \beta_{\tsub{2}} [ T_{\tsub{2}}(\zerom,t) - T^{\tsup{i}} ] \simeq 0
   \label{mfc-hl-lts-s-pg}
\end{equation}
We have cast the mathematical problem [\REqs{ic}{mfc-hl-lts-s-pg}] in non-dimensional form by selecting the solid-block half length $a$, the top-tem\-per\-a\-ture transient time $\ttd$ and the initial temperature $T^{i}$ as reference parameters for, respectively, coordinate $y$, time $t$ and temperatures $T_{k}$.
The characteristic numbers generated by this choice are tabulated in \Rta{cn} together with the selected computational cases. 
We have set the diffusion numbers $\;\rho_{\tsub{k}} c_{\tsub{k}} a^2 / \tc{k} \ttd\;$ to unity
for simplicity and have chosen the top-temperature increase as 10\% of the initial temperature
to be somehow consistent with the assumed validity of thermal-stress linearisation [\REq{hl-lts}].
\begin{table}[h]
   \renewcommand{\arraystretch}{1.5}
   \caption{Computational cases and selected values of characteristic numbers}
   \begin{tabular*}{\columnwidth}{ @{\extracolsep{\fill}} c c c c c }
      \hline
      case & $\rho_{\tsub{k}} c_{\tsub{k}} a^2 / \tc{\tsub{k}} \ttd$  & $T^{\tsup{f}} / T^{\tsup{i}}$ & $\tc{2}/\tc{1} $ & $\beta_{\tsub{2}}/\beta_{\tsub{1}}$ \\ 
      \hline
      1    & 1 & 1.1 & 1.0                        & 1.0 \\ 
      2    & 1 & 1.1 & 0.5                        & 1.0 \\ 
      3    & 1 & 1.1 & 0.5                        & 0.9 \\ 
      \hline
   \end{tabular*}
   \label{cn}
\end{table}
We have solved the mathematical non-dimensional problem numerically via two independently thought, time-accurate, finite-difference based algorithms (DG \& PSL) independently implemented in fortran codes. 
The one-dimensional grid on the $y/a$ axis comprises 102 points with a non-dimensional step of $2 \cdot 10^{-2}$; the non-dimensional time step is $5 \cdot 10^{-5}$.
The temperature-profile temporal evolutions to steady state for each computational case listed in \Rta{cn} are shown in \Rfis{pg-c1T}{pg-c3T}.
The blue, red and green profiles corresponds to, respectively, the initial situation, the end of the top-tem\-per\-a\-ture transient and the steady state.
The time necessary to attain steady state is about twice the top-temperature transient duration. 
Case 1 (\Rfi{pg-c1T}) reflects the standard situation with temperature and heat-flux continuity at the separation surface because the solids are made of same material; this case has been considered mainly to validate prediction capability of and agreement between the two independent numerical codes.
\begin{figure}[h]
  \fbox{\includegraphics[bb = 40 132 536 590 , clip , width=.975\columnwidth]{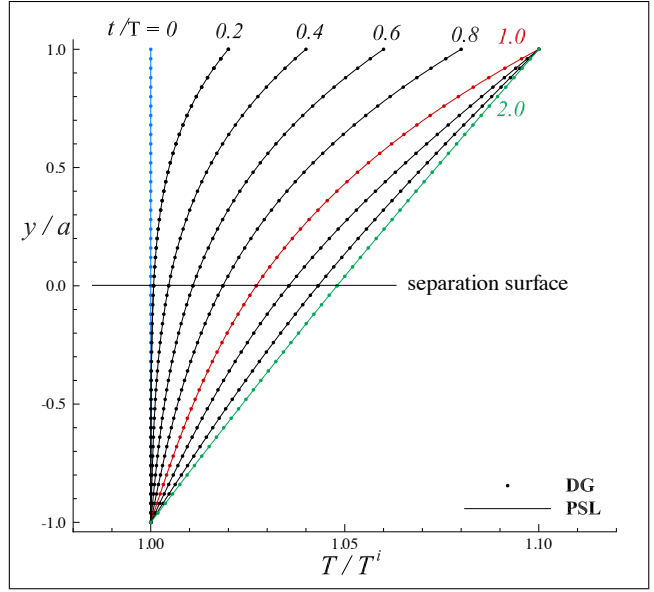}}
  \caption{Computational case 1. Solids are made by same material.}
  \label{pg-c1T}
\end{figure}
Case 2 (\Rfi{pg-c2T}) has been selected \textit{to reproduce} the standard textbook result based on the \textit{intuitive} imposition of temperature continuity [\REq{Tc}]. 
The solids are made of different materials with different thermal behaviours ($\tc{}_{2}/\tc{}_{1}=0.5$) but differences in their stress response to thermal field are ignored by forcing $\;\beta_{\tsub{2}}/\beta_{\tsub{1}}=1\;$; consequently, temperature continuity with profile-slope discontinuity are recovered. 
This approximation is, therefore, meaningful only when the thermal-stress coefficients $\beta_{\tsub{k}}$ are approximately equal.
\begin{figure}[h]
  \fbox{\includegraphics[bb = 40 150 536 620 , clip , width=.975\columnwidth]{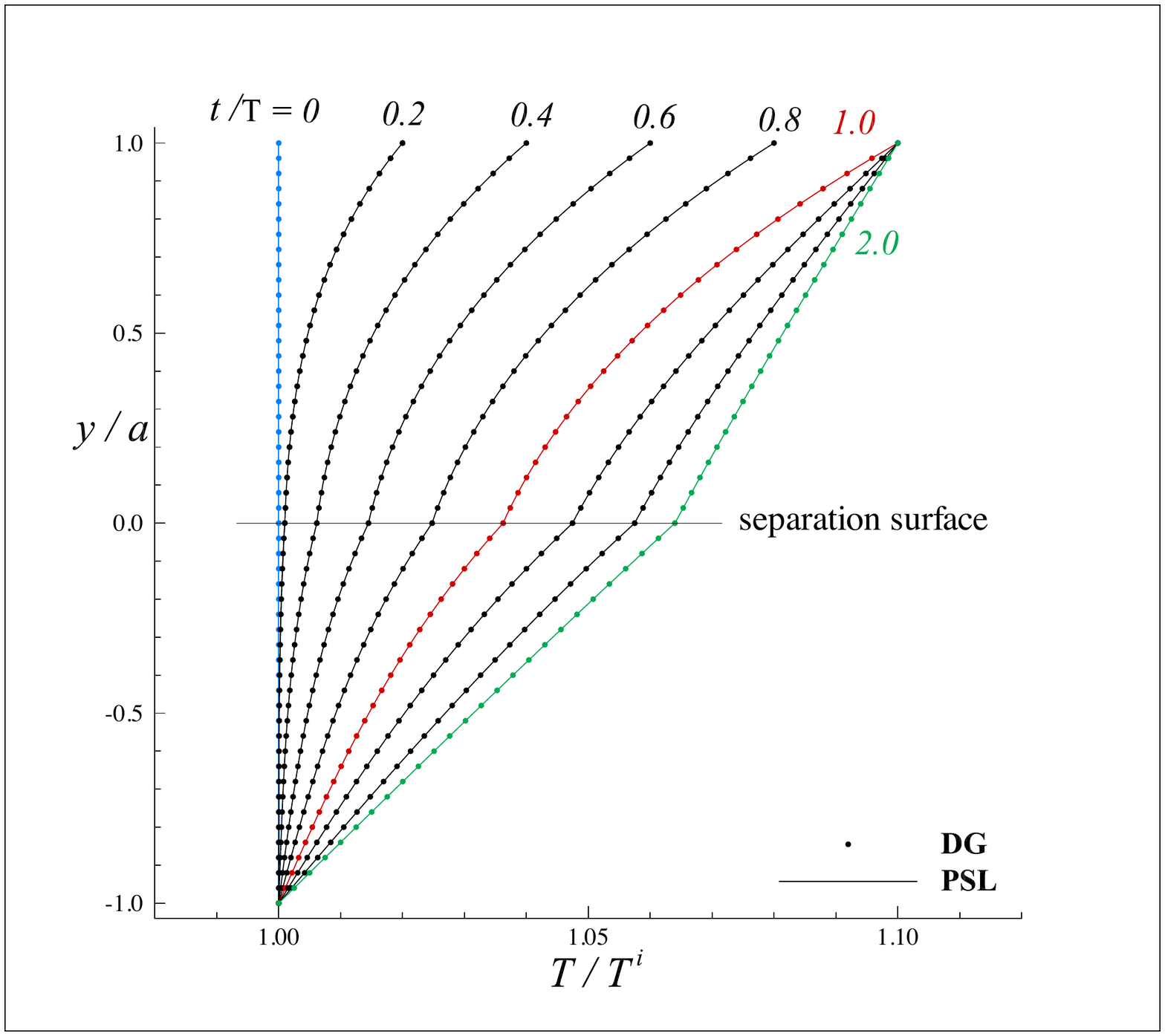}}
  \caption{Computational case 2. Solids are made by different material but differences in thermal-stress behaviour are ignored; temperature continuity is imposed (forced) at separation surface.}
  \label{pg-c2T}
\end{figure}
Case 3 (\Rfi{pg-c3T}) corresponds to the more physical situation in which heat-flux [\REqd{hfc}{hfc-fl-pg}] and tension [\REqd{mfc}{mfc-hl-lts-s-pg}] continuity prevail simultaneously at the separation surface; in this way, the materials' different stress responses to thermal field are taken into due account and, obviously, a temperature jump must settle in at the separation surface in accordance with \REq{mfc-hl-lts-s-pg} in order to secure tension continuity.
The temperature jump at the separation surface shown in \Rfi{pg-c3T} may appear surprising to minds accustomed, by consuetude in years of practice, to the profile of \Rfi{pg-c2T}; yet it is an effect guaranteed by the principle of momentum conservation as physically legitimate as the temperature-profile slope jump is guaranteed by the principle of energy conservation.
\begin{figure}[h]
  \fbox{\includegraphics[bb = 40 150 536 620 , clip , width=.975\columnwidth]{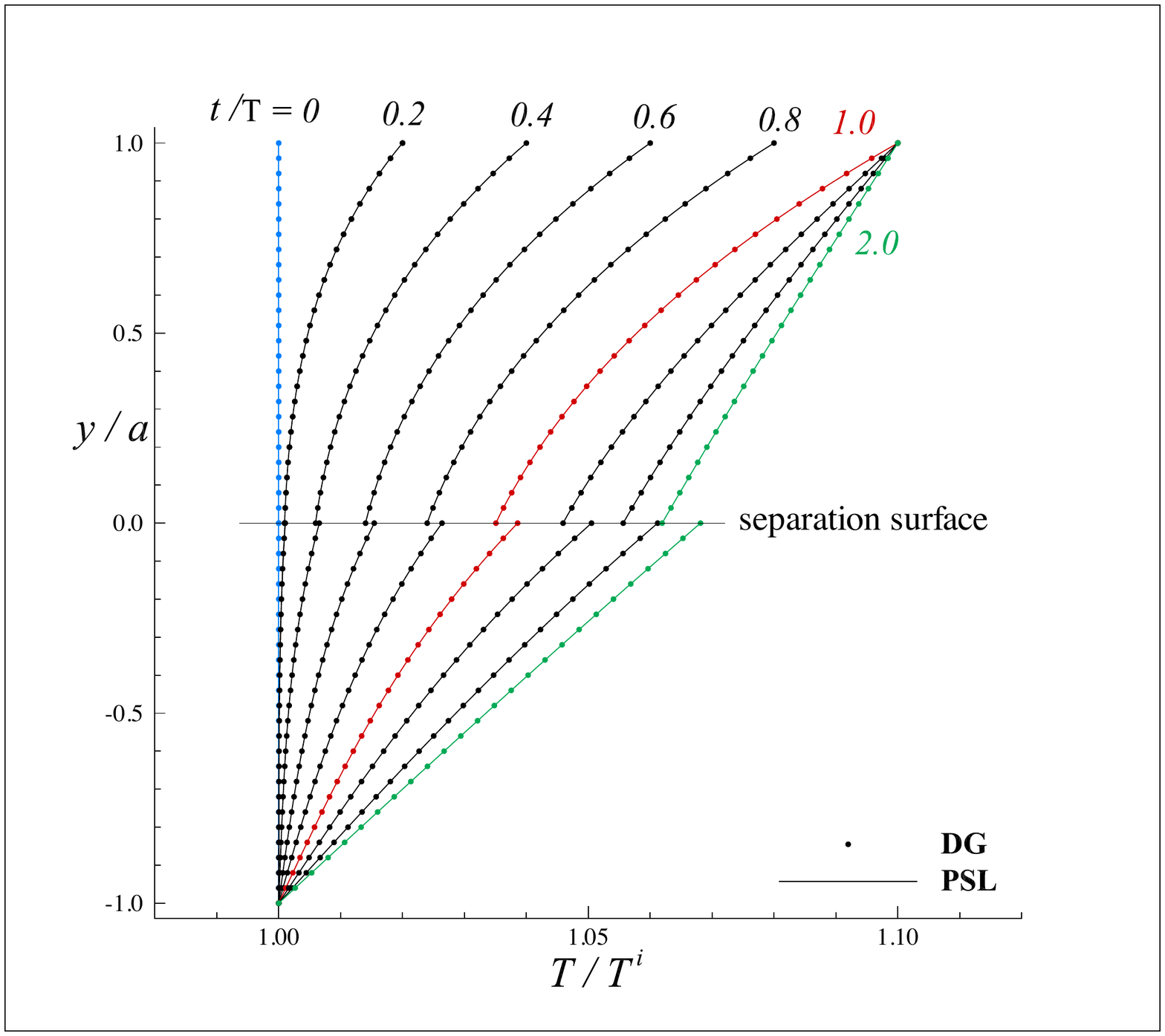}}
  \caption{Computational case 3. Solids are made by different material and differences in thermal-stress behaviour are taken into account; temperature jump at separation surface is necessary to secure tension continuity.}
  \label{pg-c3T}
\end{figure}


\section{Heat-transfer test case with interface} \label{httc-int}

\subsection{Theoretical considerations}
In the presence of an interface (\Rfis{Mvmsci}{Mvmsci.c}), the surface balance equations for mass, thermodynamic energy and momentum \cite{ln1979aa} for solids ($\V_{k}=\Vt=\W=0$) read
\begin{equation}
   \pder{\rta}{t} = 0
   \label{sbe.m}
\end{equation}
\begin{equation}
   \rta\pder{\uta}{t} - \Dives[\nuv{}\times(\nuv{}\times\dft{U})] = \nuv{1}\cdot\df{U,1}+\nuv{2}\cdot\df{U,2}
   \label{sbe.te}
\end{equation}
\begin{equation}
   \Dives[\nuv{}\times(\nuv{}\times\stt)]= -\nuv{1}\cdot\st{1} - \nuv{2}\cdot\st{2}
   \label{sbe.mom}
\end{equation}
In \REqs{sbe.m}{sbe.te}
\begin{tabbing}
  xx \= xxxxx \= xxxxxxxxxx \kill
  \> $\rta$        \> superficial mass density \\
  \> $\uta$       \> superficial thermodynamic-energy density \\
  \> $\stt$        \> superficial stress tensor \\
  \> $\dft{U}$    \> superficial thermodynamic-energy diffusive flux
\end{tabbing}
\REqb{sbe.m} enforces the constancy of the superficial mass density, a consistent analog of what happens in the solids.
\REqdb{sbe.te}{sbe.mom} stipulate unambiguously the lack of heat-flux and continuity tension. 
The heat-flux jump [right-hand side of \REq{sbe.te}] provides the thermodynamic energy deposited per unit time in the interface's unit area which, in turn, is partly (algebraically) stored locally and partly diffused \textit{tangentially} along the interface.
The tension jump [right-hand side of \REq{sbe.mom}] balances the superficial force arising from the tensional state existing inside the interface.

For consistency with the situation without interface, dealt with in \Rse{httc-std}, we retain same phenomenological relations [\REqd{hf-fl}{hl-lts}] for the solids.
For the solid interface, we assume a Fourier-like superficial thermodynamic-energy diffusive flux \cite{ln1979aa}
\begin{equation}
  \nuv{}\times(\nuv{}\times\dft{U}) = \tct \;\Grads \Tt
  \label{shf-fl}
\end{equation}
and a Hooke-like isotropic superficial stress tensor
\begin{equation} 
  \nuv{}\times(\nuv{}\times\stt) \simeq \tsct (\Tt-\Tt_{\tsub{0}})\;\uts  
  \label{shl-lts}
\end{equation}
with neglected elastic contribution and linearised thermal stress for consistency with \REq{hl-lts}.
In \REqd{shf-fl}{shl-lts}
\begin{tabbing}
  xx \= xxxxx \= xxxxxxxxxx \kill	
  \> $\Tt$       \> superficial temperature \\
  \> $\Tt_{\tsub{0}}$       \> superficial reference temperature \\
  \> $\tct$        \> superficial thermal conductivity  (tangential)\\
  \> $\tsct $    \> superficial thermal-stress coefficient (tangential) \\
  \> $\uts$           \> surface unit tensor 
\end{tabbing}
Superficial thermal conductivity and thermal-stress coefficient are considered constant.
Moreover, in line with the for-convenience assumed negligibility of deformations in both solids and solid interface, we retain for the superficial ther\-mo\-dy\-nam\-ic-energy density only the superficial-temperature dependence and assume constant the superficial constant-strain specific heat \cite{hc1963}
\begin{equation} 
  \uta(\Tt) = \uta(\Tt_{\tsub{0}}) + \tilde{c} (\Tt-\Tt_{\tsub{0}})
  \label{ste-l}
\end{equation}

On the basis of the described set of assumptions, the surface balance equations of thermodynamic energy and momentum [\REqd{sbe.te}{sbe.mom}] acquire the closed form
\begin{equation}
   \rta\tilde{c}\pder{\Tt}{t} - \tct \Laps\Tt = - \tc{1} \nuv{1} \cdot \Grad T_{\tsub{1}} - \tc{2} \nuv{2}\cdot \Grad T_{\tsub{2}}
   \label{sbe.te.cf}
\end{equation}
\begin{multline}
   \Dives[\tsct  (\Tt-\Tt_{\tsub{0}})\;\uts] \simeq \\ 
   \beta_{\tsub{1}} ( T_{\tsub{1}} - T_{\tsub{0,1}} )_{\tsub{n=\zerop}} \nuv{1} + \beta_{\tsub{2}} ( T_{\tsub{2}} - T_{\tsub{0,2}} )_{\tsub{n=\zerom}} \nuv{2}
   \label{sbe.mom.cf}
\end{multline}
The terms on the right-hand side of \REqd{sbe.te.cf}{sbe.mom.cf} are intended evaluated at the generic point $P$ of the interface (see \Rfid{Mvmsci}{Mvmsci.c}).

\subsection{Planar configuration}

\subsubsection{Differential equations and intial/boundary conditions} \label{deibc}
With reference to \Rfi{Mvmsci}, we complement the still applicable initial and boundary conditions [\REqs{avw}{bottomT}] and the heat-transfer equations [\REq{htge}] considered in \Rse{std-pg} with the following:
\begin{itemize}
     \item adiabatic vertical walls
            \begin{equation}  \Pder{\Tt}{x}{x=\pm a} = \Pder{\Tt}{z}{z=\pm a} = 0 \label{avw.i} \end{equation}
     \item uniform initial temperature
            \begin{equation} \Tt(x,z,0) = T^{\tsup{i}} \label{ic.i} \end{equation}
\end{itemize}
The surface balance equations [\REqd{sbe.te.cf}{sbe.mom.cf}] become
\begin{equation}
   \rta\tilde{c}\pder{\Tt}{t} - \tct \Laps\Tt =  \tc{1} \Pder{T_{\tsub{1}}}{y}{y=\zerop} - \tc{2} \Pder{T_{\tsub{2}}}{y}{y=\zerom} 
   \label{sbe.te.cf.p}
\end{equation}
\begin{equation}
   \tsct   \Grads \Tt \simeq - \beta_{\tsub{1}} [ T_{\tsub{1}}(\zerop,t) - T^{\tsup{i}} ] \bv{j} + \beta_{\tsub{2}} [ T_{\tsub{2}}(\zerom,t) - T^{\tsup{i}} ] \bv{j}
   \label{sbe.mom.cf.p}
\end{equation}
The projection of \REq{sbe.mom.cf.p} in the $x$ and $z$ directions yields
\begin{equation}
   \pder{\Tt}{x} = \pder{\Tt}{z} \simeq 0
   \label{sbe.mom.cf.p.sg.s}
\end{equation}
Thus, within the current approximation, the superficial temperature must be uniform along the interface, its surface gradient vanishes
and so does its surface Laplacian operator
\begin{equation}
   \Laps\Tt = \Dives\Grads \Tt = \psde{\Tt}{x}+\psde{\Tt}{z}  \simeq 0
   \label{sbe.mom.cf.p.sd}
\end{equation}
In this case, therefore, the tangential diffusion of superficial thermodynamic energy is switched off.
With these simplifications, \REqd{sbe.te.cf.p}{sbe.mom.cf.p} can be reduced to the final form
\begin{equation}
   \rta\tilde{c}\pder{\Tt}{t} = \tc{1} \Pder{T_{\tsub{1}}}{y}{y=\zerop} - \tc{2} \Pder{T_{\tsub{2}}}{y}{y=\zerom} 
   \label{sbe.te.cf.p.fin}
\end{equation}
\begin{equation}
   \beta_{\tsub{1}} [ T_{\tsub{1}}(\zerop,t) - T^{\tsup{i}} ] - \beta_{\tsub{2}} [ T_{\tsub{2}}(\zerom,t) - T^{\tsup{i}} ] \simeq 0
   \label{sbe.mom.cf.p.fin}
\end{equation}
\REqb{sbe.mom.cf.p.fin} is the same as \REq{mfc-hl-lts-s-pg} and indicates that ther\-mal-stress continuity prevails also in the presence of a planar interface because the tensional state in the interface does not generate any stress in the normal direction; this is a consequence of the absence of interface curvature, as it will appear clear in \Rse{deibc-c}.

The comparison of \REqd{sbe.te.cf.p.fin}{sbe.mom.cf.p.fin} with \REqd{hfc-fl-pg}{mfc-hl-lts-s-pg} clearly reveals that the mathematical problem we have built so far in the presence of the interface is underdetermined: one equation is missing because the number (two) of interface conditions is the same as in the case of a separation surface but \REq{sbe.te.cf.p.fin} contains the time derivative of the superficial temperature, a variable that belongs to the set of unknowns of the mathematical problem.
The occurrence of such a situation is not new. 
A quick check in the literature reveals that differential/algebraic equations are somehow assumed or derived in addition to the surface balance equations. 
In the case of fluid interfaces, for example, Waldmann \cite{lw1967zfn} selected
\begin{equation}
   \frac{1}{\Tt} = \frac{1}{2} \left( \frac{1}{T_{\tsub{1}}} + \frac{1}{T_{\tsub{2}}} \right)
\end{equation}
Bedeaux and co-authors \cite{db1976p} used arguments of LIT to derive an additional differential equation [Eq. (5.14) at page 454] for the superficial temperature that reduces to the algebraic form
\begin{equation}
   \Tt = \frac{1}{2} \left( T_{\tsub{1}} + T_{\tsub{2}} \right) \label{Ttdb1976p}
\end{equation}
in a zero-order approximation. 
{\color{black} These authors derive an interesting surface entropy-production expression [Eq. (4.16) at page 451] that contains the normal heat fluxes of the volume phases. 
Obviously, phenomenological equations for the latter terms are provided by the LIT analysis of the entropy productions in the \textit{volume} phases; instead, and rather peculiarly, the authors seemingly ignore this fact and proceed to derive from the \textit{surface} entropy production, always according to the LIT prescript, additional phenomenological equations for the terms in question, their belonging to the volume phases notwithstanding. A similar approach appears to have been followed also by Sagis \cite{ls2011rmp} within the formalism of extended irreversible thermodynamics. If Bedeaux and co-authors would have looked at their surface-entropy production from the viewpoint that Napolitano \cite{ln1979aa} exploited to prove the coincidence of the normal components of, respectively, the superficial mass velocity and the interface geometrical velocity ($\nuv{}\cdot\Vt = \nuv{}\cdot\W$) then they would have retrieved for the interface they considered the condition of temperature continuity $\Tt=T_{\tsub{1}}=T_{\tsub{2}}$. }
Napolitano himself dealt with \textit{pure} interfaces \cite{ln1978aa} under the assumption that (Sec. 2 at page 567)
\begin{quote}
   the interface ... is considered a stream surface across which the velocity and the temperature are continuous.
\end{quote}
that is equivalent to complement the surface balance equations with additional algebraic equations.
This important aspect of MTGSI is still an open issue dealt with in different manners and a final word converting workers in the field to unanimous consensus seems to have not been spoken yet.
We are aware that the resolution of the hindrance caused by such an aspect is crucial, essential and mandatory for a satisfactory establishment of MTGSI but here we will not deal with it because it is not critical for the modest purposes of this work.
In \Rse{dme} we will derive the missing equation following a conceptual pathway whose end products are consistent (to our minds) with the surface balance equations dealt with so far
[\REqd{sbe.te.cf.p.fin}{sbe.mom.cf.p.fin}]; then, we will proceed in \Rse{res.pg.i} to present and discuss numerical results.

\subsubsection{Derivation of the missing equation} \label{dme}
We begin from the configuration illustrated in \Rfi{Mvmsc3}: three solids ($k=$1,\,\tildetxt{1},2) in contact with separation surfaces as considered in \Rse{httc-std}.
For the solid in the middle ($k=\;$\tildetxt{1}), we extend a bit the phenomenological equations [\REqd{hf-fl}{hl-lts}] assumed for the other solids ($k=1,2$) : the thermal-conductivity tensor and the thermal-stress-coefficient tensor are still diagonal but not isotropic
\begin{equation}
     \df{U,\tildesub} = - (\tc{\tildesub s}\uts + \tc{\tildesub n}\ntilde\ntilde) \cdot \Grad T_{\tildesub}
     \label{hf-fl.tilde}
\end{equation}
\begin{equation}
     \st{\tildesub} \simeq  - ( \beta_{\tildesub\tsub{s}} \uts + \beta_{\tildesub\tsub{n}}\ntilde\ntilde ) ( T_{\tildesub} - T^{\tsup{i}} )  
     \label{hl-lts.tilde}
\end{equation}
The symbol \tildetxt{1} appears as subscript for the time being in analogy with the other two volume phases.
This position is meant to single out diffusion processes in the normal direction with respect to those in the tangential directions and seems legitimate in view of the forthcoming passage to the limit $\epsilon\rightarrow 0$.
Heat-flux continuity  
\begin{equation}
   \tc{1} \nuv{1} \cdot \Grad T_{\tsub{1}}|_{\tsub{+\epsilon,t}} + \tc{\tildesub\tsub{n}} \ntilde \cdot \Grad T_{\tsub{\tildesub}}|_{\tsub{+\epsilon,t}} = 0   
   \label{hfc1}
\end{equation}
\begin{equation}
   \tc{2} \nuv{2} \cdot \Grad T_{\tsub{2}}|_{\tsub{-\epsilon,t}} - \tc{\tildesub\tsub{n}}  \ntilde \cdot \Grad T_{\tsub{\tildesub}}|_{\tsub{-\epsilon,t}} = 0   
   \label{hfc2}
\end{equation}
and tension continuity 
\begin{equation}
  \beta_{\tsub{1}}  [T_{\tsub{1}}(+\epsilon,t) - T^{\tsup{i}}] \nuv{1} \cdot\ut + \beta_{\tildesub\tsub{n}} [T_{\tildesub}(+\epsilon,t) - T^{\tsup{i}} ] \ntilde \cdot  \ut= 0   
   \label{mfc1}
\end{equation}
\begin{equation}
  \beta_{\tsub{2}} [ T_{\tsub{2}}(-\epsilon,t) - T^{\tsup{i}}  ]\nuv{2} \cdot \ut - \beta_{\tildesub\tsub{n}}  [ T_{\tildesub}(-\epsilon,t) - T^{\tsup{i}} ] \ntilde \cdot  \ut = 0   
   \label{mfc2}
\end{equation}
apply at the points of the upper/lower separation surfaces.
\begin{figure}
  \fbox{\includegraphics[bb = 30 120 585 675 , clip , width=.975\columnwidth]{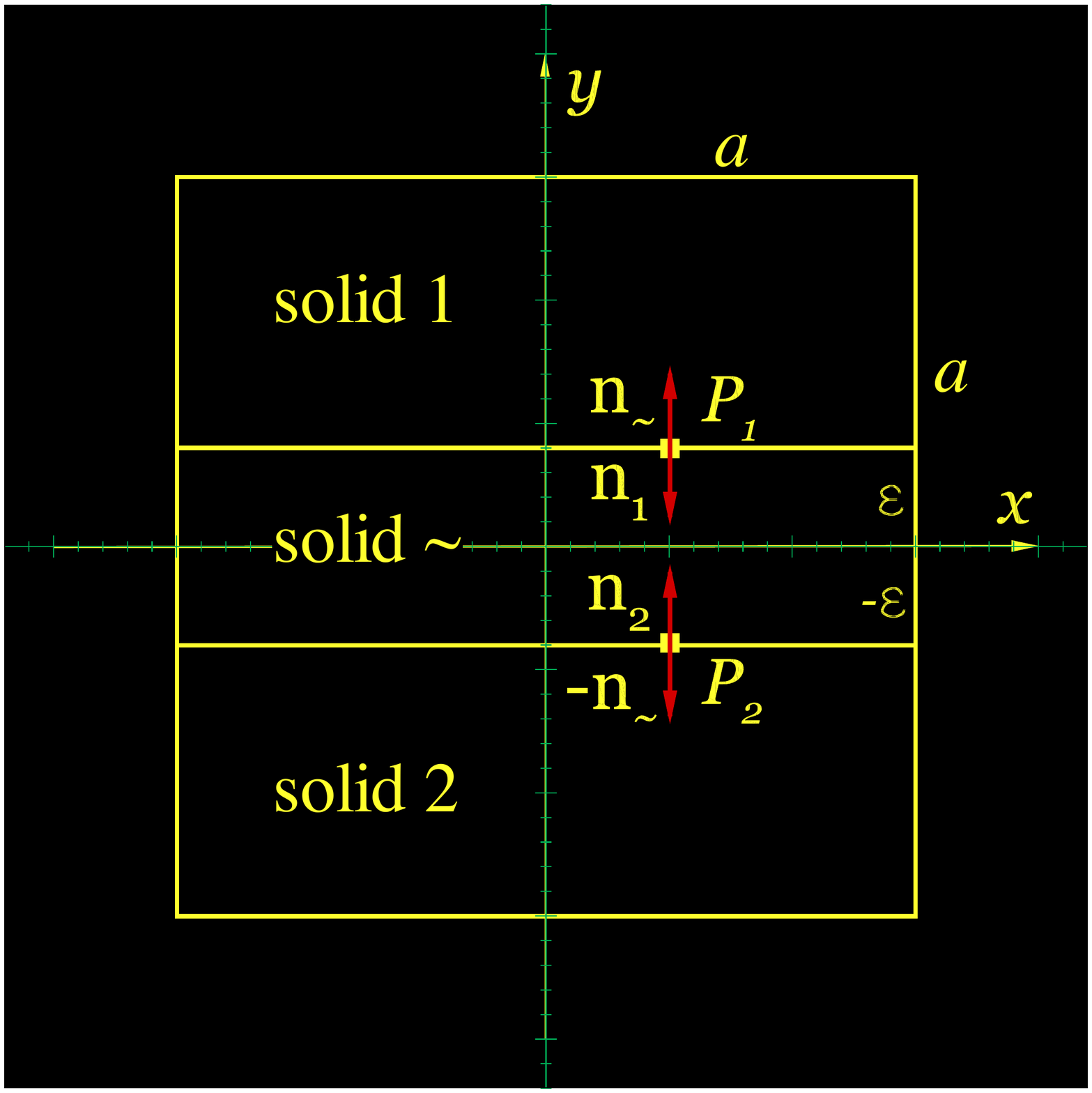}}  
  \caption{Alternative macroscopic representation of the silicon-platinum contact sketched in \Rfi{mvmsci} when $\epsilon\rightarrow 0$}
  \label{Mvmsc3}
\end{figure}
In planar configuration, \REqs{hfc1}{mfc2} can be rephrased more explicitly as 
\begin{equation}
   \tc{1} \Pder{T_{\tsub{1}}}{y}{y=+\epsilon} = \tc{\tildesub\tsub{n}} \Pder{T_{\tsub{\tildesub}}}{y}{y=+\epsilon} 
   \label{hfc1.pc}
\end{equation}
\begin{equation}
   \tc{2} \Pder{T_{\tsub{2}}}{y}{y=-\epsilon} = \tc{\tildesub\tsub{n}}  \Pder{T_{\tsub{\tildesub}}}{y}{y=-\epsilon}   
   \label{hfc2.pc}
\end{equation}
and
\begin{equation}
  \beta_{\tsub{1}}  [T_{\tsub{1}}(+\epsilon,t) - T^{\tsup{i}}] - \beta_{\tildesub\tsub{n}} [T_{\tildesub}(+\epsilon,t) - T^{\tsup{i}} ] = 0   
   \label{mfc1.pc}
\end{equation}
\begin{equation}
  \beta_{\tsub{2}} [ T_{\tsub{2}}(-\epsilon,t) - T^{\tsup{i}}  ] - \beta_{\tildesub\tsub{n}}  [ T_{\tildesub}(-\epsilon,t) - T^{\tsup{i}} ] = 0   
   \label{mfc2.pc}
\end{equation}
The heat transfer through the solid $\tildetxt{1}$ is governed by the differential equation
\begin{equation}
  \rho_{\tildesub} c_{\tildesub} \pder{T_{\tildesub}}{t} = \tc{\tildesub\tsub{n}}\psde{T_{\tildesub}}{y} 
  \label{htge.stilde}
\end{equation}
The integration of \REq{htge.stilde} along $y$ from $-\epsilon$ to $+\epsilon$ yields
\begin{equation}
  \rho_{\tildesub} c_{\tildesub} \int\limits_{-\epsilon}^{+\epsilon} \pder{T_{\tildesub}}{t}dy = \tc{\tildesub\tsub{n}}\int\limits_{-\epsilon}^{+\epsilon} \psde{T_{\tildesub}}{y} dy 
  \label{htge.stilde.i}
\end{equation}
The time derivative commutes with the integral and the term on the right-hand side is easily integrated
\begin{multline}
   \rho_{\tildesub} c_{\tildesub} \pder{}{t} \int\limits_{-\epsilon}^{+\epsilon} T_{\tildesub} dy = 
   \tc{\tildesub\tsub{n}} \Pder{T_{\tsub{\tildesub}}}{y}{y=+\epsilon} - \tc{\tildesub\tsub{n}} \Pder{T_{\tsub{\tildesub}}}{y}{y=-\epsilon}
  \label{htge.stilde.i.p}
\end{multline}
We introduce the average temperature
\begin{equation}
   \aTt = \frac{1}{2 \epsilon} \int\limits_{-\epsilon}^{+\epsilon} T_{\tildesub} dy \label{aTt}
\end{equation} 
and take advantage now of heat-flux continuity [\REqd{hfc1.pc}{hfc2.pc}] on both separation surfaces to recast \REq{htge.stilde.i.p} into the equivalent form
\begin{multline}
   (2 \epsilon \rho_{\tildesub}) c_{\tildesub} \pder{\aTt}{t}  = 
   \tc{1} \Pder{T_{\tsub{1}}}{y}{y=+\epsilon} - \tc{2} \Pder{T_{\tsub{2}}}{y}{y=-\epsilon}
  \label{htge.stilde.i.p.1}
\end{multline}
which, passed to the limit $\epsilon\rightarrow 0$, assumes the structure of \REq{sbe.te.cf.p.fin} and, by comparison, leads to the following definitions
\begin{equation}
   \lim_{\epsilon\rightarrow 0}\; (2 \epsilon \rho_{\tildesub}) = \rta \label{rhot}
\end{equation}
\begin{equation}
   \lim_{\epsilon\rightarrow 0}\; c_{\tildesub} = \tilde{c} \label{sht}
\end{equation}
\begin{equation}
   \lim_{\epsilon\rightarrow 0}\; \aTt = \Tt \label{Tt}
\end{equation}
{\color{black} The existence of the limits defined by \REqs{rhot}{Tt} may seem a strong assumption but, on the other hand, its justification is prompted by the fact that it sanctions the mathematical equivalence between \REqd{sbe.te.cf.p.fin}{htge.stilde.i.p.1}.}
Next, we turn to the tension-continuity boundary conditions [\REqd{mfc1.pc}{mfc2.pc}]: mathematically speaking, they are two independent equations and
we can recast them by subtraction/addition into the equivalent, and obviously still independent, forms
\begin{multline}
  \beta_{\tildesub\tsub{n}} [T_{\tildesub}(+\epsilon,t) - T^{\tsup{i}} ] - \beta_{\tildesub\tsub{n}}  [ T_{\tildesub}(-\epsilon,t) - T^{\tsup{i}} ] = \\
  \beta_{\tsub{1}}  [T_{\tsub{1}}(+\epsilon,t) - T^{\tsup{i}}] - \beta_{\tsub{2}} [ T_{\tsub{2}}(-\epsilon,t) - T^{\tsup{i}}  ]   
  \label{mfc-.pc}
\end{multline}
\begin{multline}
  \beta_{\tildesub\tsub{n}} [T_{\tildesub}(+\epsilon,t) - T^{\tsup{i}} ] + \beta_{\tildesub\tsub{n}}  [ T_{\tildesub}(-\epsilon,t) - T^{\tsup{i}} ] = \\
  \beta_{\tsub{1}} [T_{\tsub{1}}(+\epsilon,t) - T^{\tsup{i}} ] + \beta_{\tsub{2}} [ T_{\tsub{2}}(-\epsilon,t) - T^{\tsup{i}}  ]   
  \label{mfc+.pc}
\end{multline}
If we pass \REq{mfc-.pc} to the limit $\epsilon\rightarrow 0$ then we re-obtain \REq{sbe.mom.cf.p.fin}.
\REqb{mfc+.pc} is still exploitable and its adequate manipulation leads to the missing equation we are looking for to close our mathematical problem.
We accomplish this task by rewriting the left-hand side of \REq{mfc+.pc} as
\begin{multline}
      \beta_{\tildesub\tsub{n}} [T_{\tildesub}(+\epsilon,t) - \aTt ] + \beta_{\tildesub\tsub{n}}  [ T_{\tildesub}(-\epsilon,t) - \aTt ] \\
      \mbox{\hspace{8ex}} + 2 \beta_{\tildesub\tsub{n}} [\aTt - T^{\tsup{i}}] \\
      = \beta_{\tsub{1}} [T_{\tsub{1}}(+\epsilon,t) - T^{\tsup{i}} ] + \beta_{\tsub{2}} [ T_{\tsub{2}}(-\epsilon,t) - T^{\tsup{i}}  ]
   \label{mfc+.pc.1}
\end{multline}
 and then passing to the limit $\epsilon\rightarrow 0$. 
 In so doing, we assume reasonably that
 \begin{equation}
   \lim_{\epsilon\rightarrow 0}\; [T_{\tildesub}(\pm\epsilon,t) - \aTt ] = 0 \label{aTtdiff}
\end{equation}
we register the definition 
\begin{equation}
      \lim_{\epsilon\rightarrow 0}\; \beta_{\tildesub\tsub{n}} = \tscn \label{betatn}
\end{equation}
and consequently obtain the missing equation 
\begin{equation}
  2 \tscn ( \Tt - T^{\tsup{i}} ) = 
  \beta_{\tsub{1}} [T_{\tsub{1}}(\zerop,t) - T^{\tsup{i}} ] + \beta_{\tsub{2}} [ T_{\tsub{2}}(\zerom,t) - T^{\tsup{i}}  ]   
  \label{mfc+.pc.f}
\end{equation}
that must be associated to \REqd{sbe.te.cf.p.fin}{sbe.mom.cf.p.fin}. 
It ought to be noticed that \REq{mfc+.pc.f} reduces to the result [\REq{Ttdb1976p}] of Bedeaux and co-authors \cite{db1976p} in the particular case $\;\tscn=\beta_{\tsub{1}}=\beta_{\tsub{2}}$.
An incidental remark is due at this point. 
After having seen how useful addition between \REqd{mfc1.pc}{mfc2.pc} is for the purpose of obtaining the missing equation, the attentive reader may have noticed that, as a matter of fact, we took advantage of the subtraction between \REqd{hfc1.pc}{hfc2.pc} to obtain \REq{htge.stilde.i.p.1} and be wondering what happens instead in case of addition.
It is a legitimate observation with a straightforward answer. 
Skipping the algebra, the mentioned operation together and the subsequent passage to the limit $\epsilon\rightarrow 0$ generates the following equation 
\begin{multline}\label{hfc-add}
      \lim_{\epsilon\rightarrow 0}\left[ \tc{\tildesub\tsub{n}} \Pder{T_{\tsub{\tildesub}}}{y}{y=+\epsilon}+ \tc{\tildesub\tsub{n}} \Pder{T_{\tsub{\tildesub}}}{y}{y=-\epsilon} \right]
                                       = \\ \tc{1} \Pder{T_{\tsub{1}}}{y}{y=\zerop} + \tc{2} \Pder{T_{\tsub{2}}}{y}{y=\zerom}
\end{multline}
\REqb{hfc-add} is certainly another equations but is unhelpful because it introduces also an additional unknown, that is, the quantity on its left-hand side, which does not intervene in any other place in the set of governing equations. 
This \textit{additional} equation is, therefore, a mathematical blind alley and is not needed.

Coming back to \REq{mfc+.pc.f}, we looked at it, stepped back for a thoughtful moment and wondered about its physical meaning. 
More importantly, we asked ourselves if there is a more physically profound way to deduce it rather than our \textit{ad-hoc} manner.
This is, of course, a very interesting and captivating problem but we felt it is a bit outside the reach of our preliminary study; therefore, for the time being, we decided to stamp the problem as \textit{future work} (\Rse{concl}) and to carry on with the numerical calculations.

\subsubsection{Results}\label{res.pg.i}

With the aid of \REq{mfc+.pc.f}, our mathematical problem is thus closed and we have cast it in non-dimensional form with the same reference parameters selected in \Rse{std-pg} [see text after \REq{mfc-hl-lts-s-pg}].
Two other characteristic numbers appear in addition to those tabulated in \Rta{cn}; they contain physical variables related to the interface and are shown in the rightmost columns of \Rta{cn.i} which lists also the selected computational cases.
The temperature-profile temporal evolutions corresponding to these computational cases are illustrated in \Rfis{pg-c1T.i}{pg-c6T.i}. 
We recall the colour convention: blue, red and green profiles corresponds to, respectively, initial situation, end of the top-tem\-per\-a\-ture transient and steady state, respectively.
The time necessary to attain steady state is about four times the top-temperature transient duration, therefore twice as long the time required to achieve steady state without interface. 
The square symbols indicate the values of $\Tt/T^{\tsup{i}}$ (results from both numerical codes superpose very precisely).
Cases 1-3 are analogous to the those analysed in \Rse{std-pg} (see \Rta{cn}) and have been considered to put in evidence the effects due to the presence of the interface.
\begin{table}[h] 
   \renewcommand{\arraystretch}{1.5} \setlength{\tabcolsep}{0.5\tabcolsep}
   \caption{Computational cases and selected values of characteristic numbers}
   \begin{tabular*}{\columnwidth}{@{\extracolsep{\fill}} c c c c c c c }
      \hline
      case & $\rho_{\tsub{k}} c_{\tsub{k}} a^2 / \tc{\tsub{k}} \ttd$  & $T^{\tsup{f}} / T^{\tsup{i}}$ & $\tc{2}/\tc{1} $ & $\beta_{\tsub{2}}/\beta_{\tsub{1}}$ 
           & $\rta \tilde{c} a / \tc{1} \ttd$ & $\tscn/\beta_{\tsub{1}}$ \\ 
      \hline
      1    & 1 & 1.1 & 1.0                        & 1.0 & 1 & 1.0 \\ 
      2    & 1 & 1.1 & 0.5                        & 1.0 & 1 & 1.0 \\ 
      3    & 1 & 1.1 & 0.5                        & 0.9 & 1 & 0.9 \\ 
      4    & 1 & 1.1 & 0.5                        & 0.9 & 1 & 1.0 \\ 
      5    & 1 & 1.1 & 0.5                        & 1.1 & 1 & 0.9 \\ 
      6    & 1 & 1.1 & 1.0                        & 1.0 & 1 & \textcolor{black}{$\ll 1$}\\ 
      \hline
   \end{tabular*}
   \label{cn.i}
\end{table}
In case 1 (\Rfi{pg-c1T.i}), the temperature-profile slopes at the interface show a clear difference, notwithstanding that $\tc{2}/\tc{1}=1$, in accordance with \REq{sbe.te.cf.p.fin}: the heat-flux difference (right-hand side) feeds into the superficial thermodynamic energy (left-hand side) and slope continuity is recovered only with the steady-state attainment. 
The continuity of \textit{all} temperatures at the interface is a consequence of the assumption that the interface stress response to thermal field is similar to the one ($\beta_{\tsub{2}}/\beta_{\tsub{1}}=\tscn/\beta_{\tsub{1}}=1$) of the material composing the solids 1 and 2. 
It constitutes a further evidence of validation for our numerical codes because it can be verified analytically from \REqd{sbe.mom.cf.p.fin}{mfc+.pc.f}.
\begin{figure}[h]
  \fbox{\includegraphics[bb = 40 156 540 618 , clip , width=.975\columnwidth]{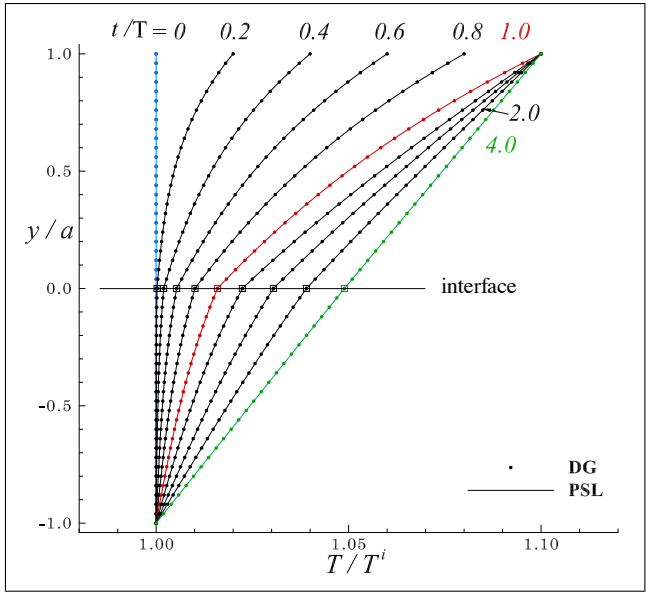}}
  \caption{Computational case 1. Solids are made by same material.}
  \label{pg-c1T.i}
\end{figure}
The temperature-profile evolutions for case 2 (\Rfi{pg-c2T.i}), case 3 (\Rfi{pg-c3T.i}) and case 4 (\Rfi{pg-c4T.i}) show consistent behaviour.
The interface temperature in cases 3 and 4 coincides with, respectively, $T_{\tsub{2}}(\zerop,t)$ and $T_{\tsub{1}}(\zerop,t)$ because $\tscn=\beta_{\tsub{2}}$ in the
former case and $\tscn=\beta_{\tsub{1}}$ in the latter case.
\begin{figure}[h]
  \fbox{\includegraphics[bb = 40 156 540 618 , clip , width=.975\columnwidth]{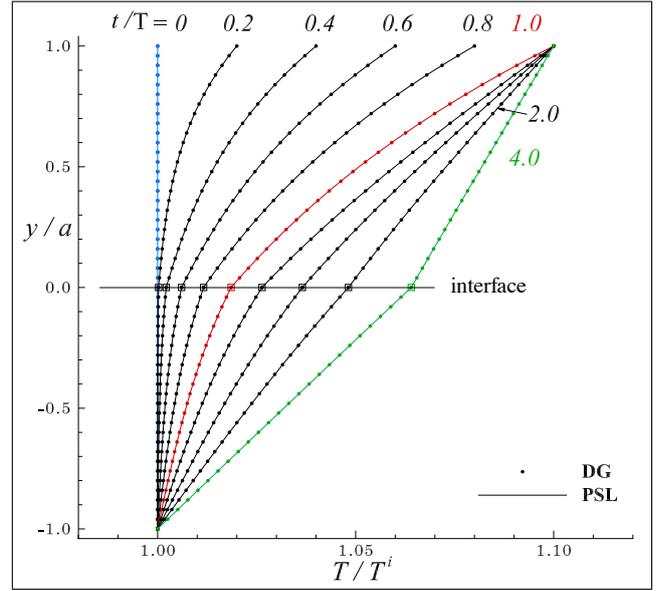}}
  \caption{Computational case 2. Solids are made by different material but differences in thermal-stress behaviour are ignored; temperature continuity is imposed (forced) at interface.  Superficial-temperature coincidence with solids' temperatures is a consequence.}
  \label{pg-c2T.i}
\end{figure}
\begin{figure}[h]
  \fbox{\includegraphics[bb = 40 156 540 618 , clip , width=.975\columnwidth]{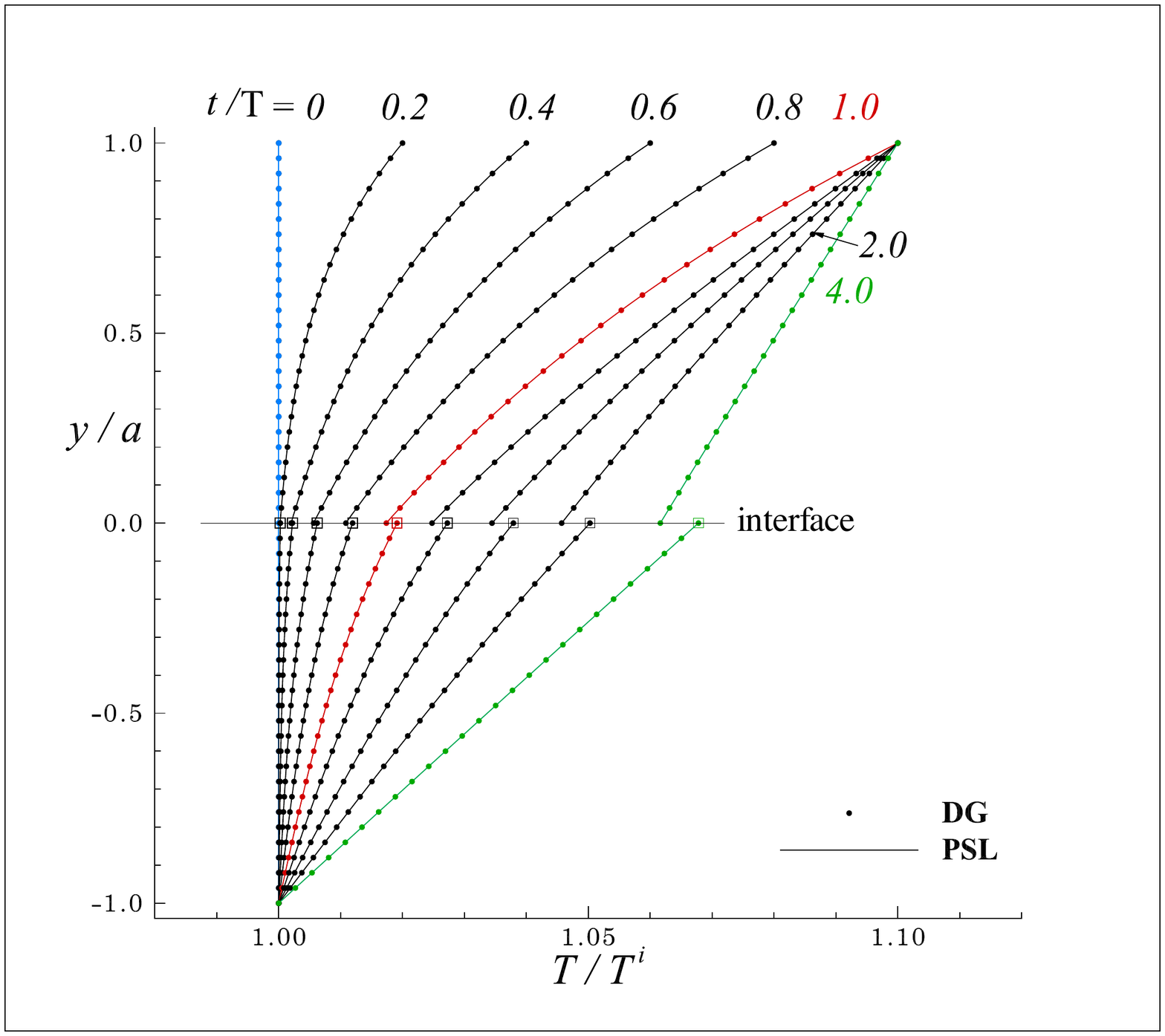}}
  \caption{Computational case 3. Solids are made by different material and differences in thermal-stress behaviour are taken into account; temperature jump at interface is necessary to secure tension continuity. Superficial temperature coincides with that of solid 2 because $\tscn=\beta_{\tsub{2}}$.}
  \label{pg-c3T.i}
\end{figure}
\begin{figure}[h]
  \fbox{\includegraphics[bb = 40 156 540 618 , clip , width=.975\columnwidth]{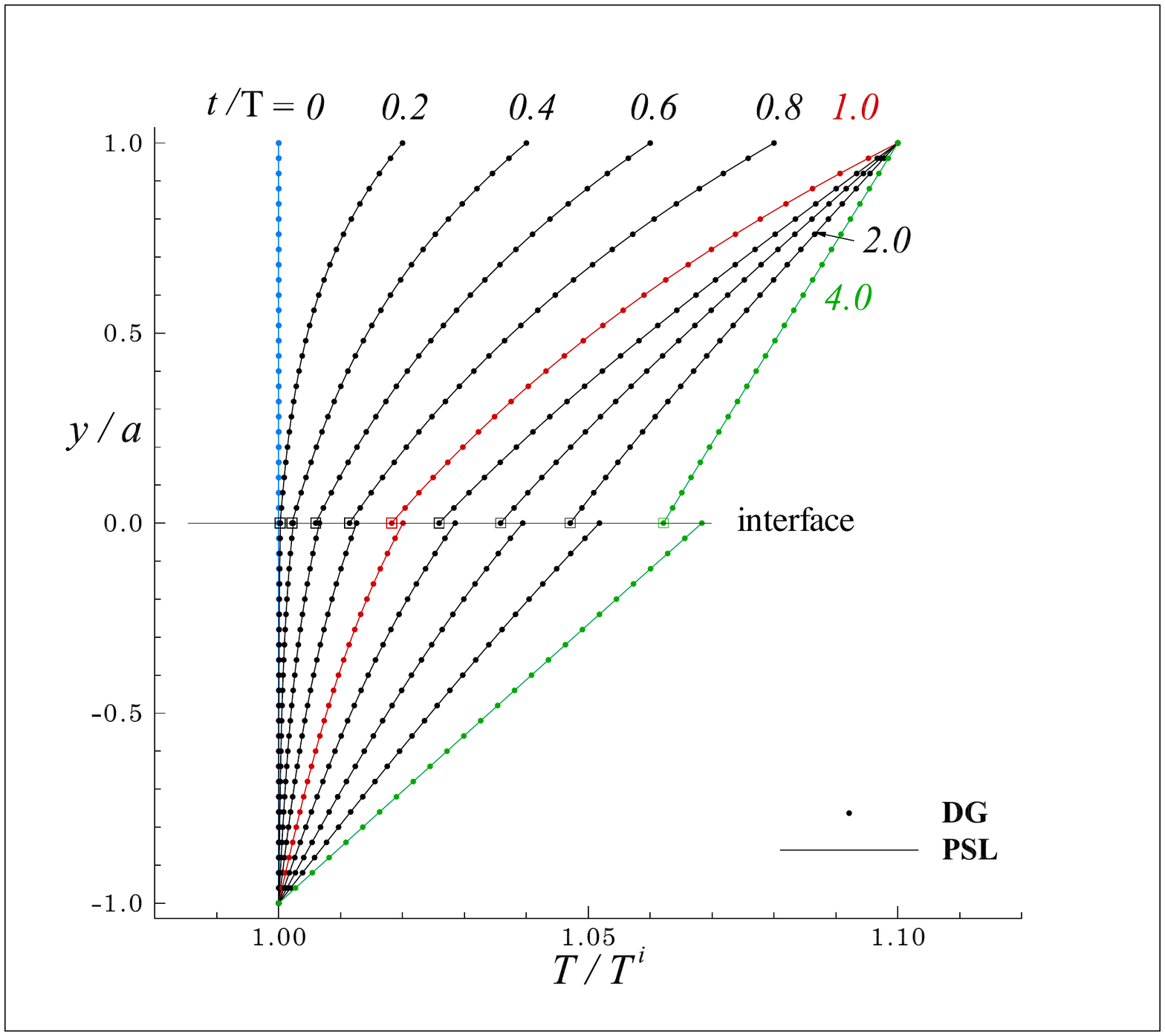}}
  \caption{Computational case 4. Similar to case 3 but superficial temperature coincides with that of solid 1 because $\tscn=\beta_{\tsub{1}}$.}
  \label{pg-c4T.i}
\end{figure}
Case 5 (\Rfi{pg-c5T.i}) is really interesting from a physical point of view and, at the same time, significant from an engineering point of view.
The temperature profiles in \Rfid{pg-c3T.i}{pg-c5T.i}, when compared with the temperature profiles of \Rfi{pg-c3T}, provide clear evidence of the thermal protection action of the interface on solid 2. In order to emphasise such evidence, we have collected in \Rta{etp} meaningful values of temperatures and heat fluxes at steady state ($t/\ttd\simeq 2$ for case 3$^{a}$ and $t/\ttd\simeq 4$ 
for cases 3$^{b}$ and 5) for comparison.
\begin{table}[h]
   \renewcommand{\arraystretch}{1.5}
   \setlength{\tabcolsep}{0.5\tabcolsep}
   \caption{Comparison of selected values of temperatures and heat fluxes at steady state}
   \begin{tabular*}{\columnwidth}{ @{\extracolsep{\fill}} l c c c c c }
      \hline &&&&& \\[-3ex]
      case & $T_{\tsub{1}}(\zerop) / T^{\tsup{i}}$  
           & $\displaystyle\frac{a}{T^{\tsup{i}}}\Pder{T_{\tsub{1}}}{y}{\zerop}$ 
           & $\Tt / T^{\tsup{i}} $ 
           & $T_{\tsub{2}}(\zerom) / T^{\tsup{i}}$ 
           & $\displaystyle\frac{a}{T^{\tsup{i}}}\Pder{T_{\tsub{2}}}{y}{\zerom}$  \\ 
      &&&&& \\[-3ex] \hline
      3$^{a}$    & 1.062 & 3.548$\cdot 10^{-2}$ &          & 1.068 & 7.097$\cdot 10^{-2}$  \\ 
      3$^{b}$    & 1.062 & 3.740$\cdot 10^{-2}$ & 1.068 & 1.068  & 6.886$\cdot 10^{-2}$  \\ 
      5            & 1.065 & 3.368$\cdot 10^{-2}$ & 1.072 & 1.059  & 6.029$\cdot 10^{-2}$  \\ 
      \hline
   \end{tabular*}
   \\[.25\baselineskip] $^{a}$ \mbox{\scriptsize without interface (\Rta{cn})} \\ $^{b}$ \mbox{\scriptsize with interface (\Rta{cn.i})} 
   \label{etp}
\end{table}
The presence of the interface mitigates the heat flux received by solid 2 (rightmost column in \Rta{etp}) as much as 15 \% between cases 3$^{a}$ and 5.
\begin{figure}[h]
  \fbox{\includegraphics[bb = 40 156 540 618 , clip , width=.975\columnwidth]{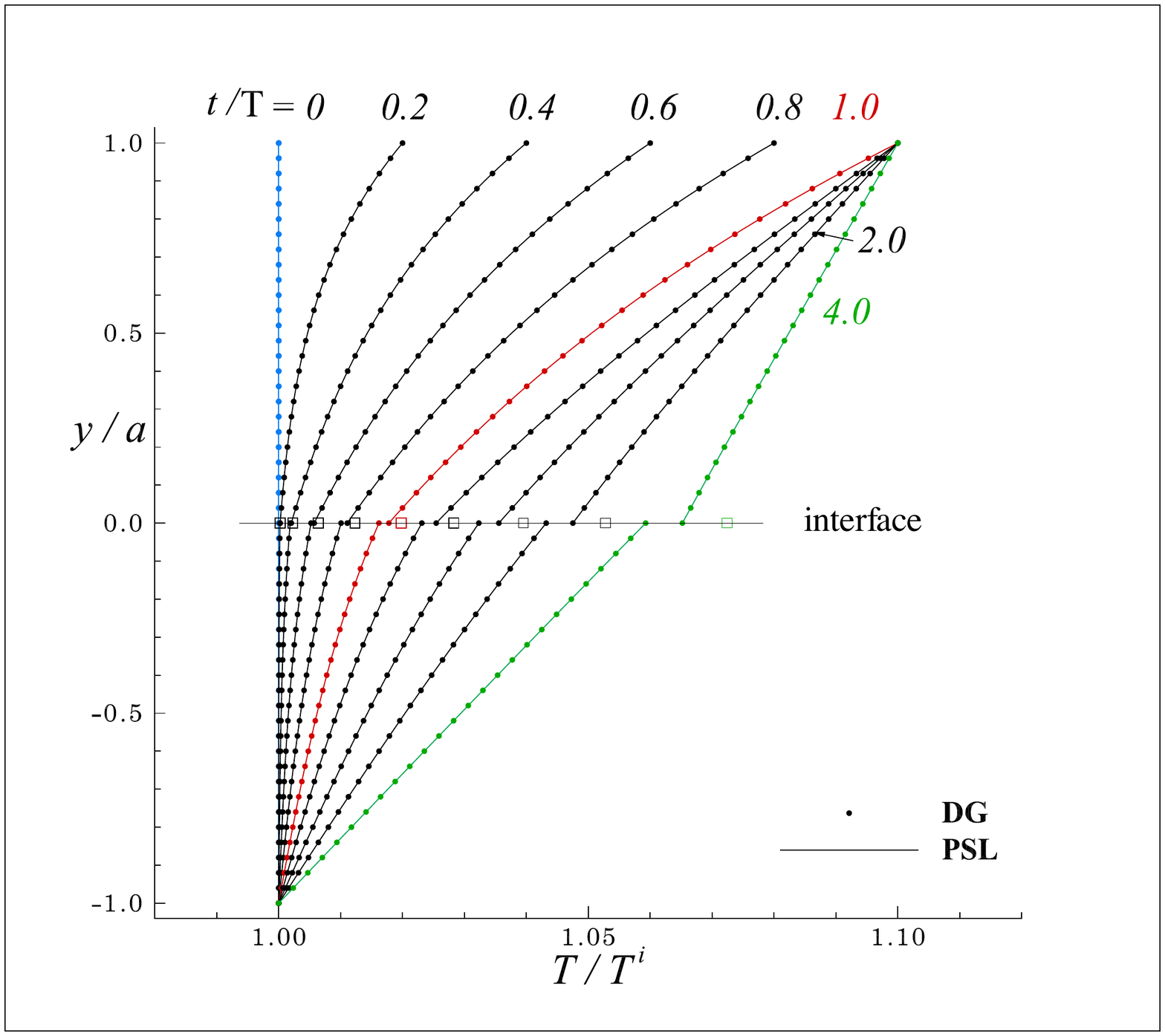}}
  \caption{Computational case 5. Solids are made by different material and differences in thermal-stress behaviour are taken into account; temperature jump at separation surface is necessary to secure tension continuity. Superficial temperature governed by heat-flux difference at interface according to \REq{sbe.te.cf.p.fin}.}
  \label{pg-c5T.i}
\end{figure}
Case 6 (\Rfi{pg-c6T.i}) represents the extreme, and therefore ideal, situation when \textcolor{black}{$\tscn\ll \beta_{\tsub{1}}=\beta_{\tsub{2}}$}.
In this case, the thermal-protection action of the interface is complete: solid 2 is totally shielded and does not receive any heat flux because the heat flux outgoing from solid 1 goes exclusively to replenish the superficial thermodynamic energy of the interface whose superficial temperature grows monotonically with time even after the attainment of the steady state ($t/\ttd\simeq 1.5$) in solid 1 (solid squares).
The results shown in \Rfi{pg-c6T.i} are analogous to and confirm those obtained by Schmidtmann \cite{bs2013vki} with a physically simpler model.
\begin{figure}[h]
  \fbox{\includegraphics[bb = 40 156 540 618 , clip , width=.975\columnwidth]{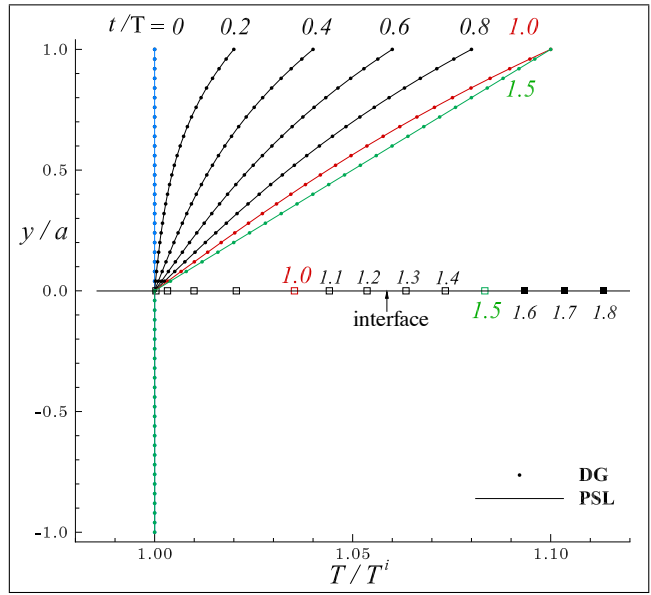}}
  \caption{Computational case 6. Ideal situation of complete thermal protection when \textcolor{black}{$\tscn\ll \beta_{\tsub{1}}=\beta_{\tsub{2}}$}, analogous to the case studied by Schmidtmann \cite{bs2013vki}.}
  \label{pg-c6T.i}
\end{figure}

\subsection{Cylindrical configuration} \label{i-cg}

\subsubsection{Differential equations and initial/boundary conditions} \label{deibc-c}
For the cylindrical configuration (see \Rfi{Mvmsci.c}), we adopt body-fitted coordinates: the curvilinear abscissa $s$ reckoned from point A and the normal distance $n$ from the interface (positive/negative in solid 1/2).
We assume
\begin{itemize}
	\item uniform initial temperature
	       \begin{equation} T_{\tsub{k}}(s,n,z,0) = \Tt(s,z,0) = T^{\tsup{i}} \label{uit} \end{equation}
	\item uniform temperature at the top of solid 1 increasing linearly in time, within a finite interval $\ttd$, from the initial value $T^{\tsup{i}}$ to a final value $T^{\tsup{f}}$ 
	\begin{equation}
			T_{\tsub{1}}(s,+R,z,t) = \left\{ 
												\begin{array}{l@{\hspace{1.5em}}c} T^{\tsup{i}} & \hspace{1.75em} t < 0 \\ 
	                                            \displaystyle  T^{\tsup{i}} + (T^{\tsup{f}}-T^{\tsup{i}})\frac{t}{\ttd} &  0\leq t \leq \ttd \\ 
	                                            T^{\tsup{f}} & \hspace{-1.75em} \ttd < t \end{array}  \right.
	\label{topTc}
	\end{equation}
\end{itemize}
These conditions determine an one-dimensional heat transfer in the radial direction governed by the standard diffusion equation
\begin{equation}
		\rho_{\tsub{k}} c_{\tsub{k}} \pder{T_{\tsub{k}}}{t} = \tc{k}\psde{T_{\tsub{k}}}{n} + \frac{\tc{k}}{n+R} \pder{T_{\tsub{k}}}{n}
		\label{htge.c}
\end{equation}
\REqb{htge.c} must be handled with care on the cylinder axis ($n=-R \; ; \; k=2$) because the scale factor in the denominator of the second term on the right-hand side vanishes.
In order to avoid the discontinuity, therefore, we must assume the ulterior boundary condition
\begin{equation} \label{ndoca}
		\Pder{T_{\tsub{2}}}{n}{n=-R} = 0
\end{equation}
Consequently, the second term on the right-hand side becomes a mathematically undetermined form resolvable by the de l'H\^{o}pital's theorem 
\begin{equation}\label{ndoca.st}
		\lim_{n \rightarrow -R} \; \frac{1}{n+R} \pder{T_{\tsub{2}}}{n} = \Psde{T_{\tsub{2}}}{n}{n=-R}
\end{equation}
and the governing equation [\REq{htge.c}] becomes discontinuity-free
\begin{equation}  \label{htge.c.df}
		\rho_{\tsub{2}} c_{\tsub{2}} \Pder{T_{\tsub{2}}}{t}{n=-R} = 2\tc{2}\Psde{T_{\tsub{2}}}{n}{n=-R} 
\end{equation}
The surface balance equations [\REqd{sbe.te.cf}{sbe.mom.cf}] become
\begin{equation}   \label{sbe.te.cf.c}
		\rta\tilde{c}\pder{\Tt}{t} - \tct \Laps\Tt =  \tc{1} \Pder{T_{\tsub{1}}}{n}{n=\zerop} - \tc{2} \Pder{T_{\tsub{2}}}{n}{n=\zerom} 
\end{equation}
\begin{multline}
        \tsct   \Grads \Tt - \frac{\tsct  }{R}  (\Tt-T^{\tsup{i}}) \ven \simeq \\ 
		- \beta_{\tsub{1}} [ T_{\tsub{1}}(\zerop,t) - T^{\tsup{i}} ] \ven + \beta_{\tsub{2}} [ T_{\tsub{2}}(\zerom,t) - T^{\tsup{i}} ] \ven
        \label{sbe.mom.cf.c}
\end{multline}
From the comparison between \REqd{sbe.mom.cf.p}{sbe.mom.cf.c}, we notice the appearance of the interface-curvature effect on the left-hand side of \REq{sbe.mom.cf.c}: the tensional state \textit{inside} the \textit{curved} interface features a tension component \textit{outside} of the interface itself in the normal direction, an interesting although peculiar physical occurrence already encountered by Laplace during his studies on capillarity \cite{psl1805tmc} slightly more than 200 years ago, that removes the continuity of volume-phase tensions [recall text in \Rse{deibc} after \REq{sbe.mom.cf.p.fin}]. 
Once again, the projection of \REq{sbe.mom.cf.c} along tangential and axial directions informs that the superficial temperature must be uniform along the interface
\begin{equation}
  \pder{\Tt}{s} = \pder{\Tt}{z} \simeq 0
  \label{sbe.mom.cf.p.sg.s.c}
\end{equation}
so that its surface gradient and Laplacian operator vanish
\begin{equation}
  \Laps\Tt = \Dives\Grads \Tt = \psde{\Tt}{s}+\psde{\Tt}{z}  \simeq 0
  \label{sbe.mom.cf.p.sd.c}
\end{equation}
Even for curved interface, therefore, the tangential diffusion of superficial thermodynamic energy is switched off.
With these simplifications, \REqd{sbe.te.cf.c}{sbe.mom.cf.c} can be reduced to the final form
\begin{equation}   \label{sbe.te.cf.c.f}
  \rta\tilde{c}\pder{\Tt}{t} =  \tc{1} \Pder{T_{\tsub{1}}}{n}{n=\zerop} - \tc{2} \Pder{T_{\tsub{2}}}{n}{n=\zerom} 
\end{equation}
\begin{equation}
   \frac{\tsct  }{R}  (\Tt-T^{\tsup{i}}) \simeq
   \beta_{\tsub{1}} [ T_{\tsub{1}}(\zerop,t) - T^{\tsup{i}} ]  - \beta_{\tsub{2}} [ T_{\tsub{2}}(\zerom,t) - T^{\tsup{i}} ] 
   \label{sbe.mom.cf.c.f}
\end{equation}
Of course, we face now the same situation encountered with the planar interface: we need one additional equation.
For this purpose, we have exported the reasoning based on the configuration illustrated in \Rfi{Mvmsc3} to the cylindrical configuration shown in \Rfi{Mvmsci.c} and have retrieved again \REq{mfc+.pc.f}; we omit here the detailed algebra because its development follows smoothly guidelines similar to those described in \Rse{dme} for the planar interface.

The mathematical problem cast in non-dimensional form features the same characteristic numbers shown in \Rta{cn.i} but with a simple variation: the characteristic length $a$ belonging to the planar interface must be replaced with the internal-solid radius $R$. 
The curvature term on the left-hand side of \REq{sbe.mom.cf.c} generates the additional characteristic number $ \tsct / \beta_{\tsub{1}} R $. 

\subsubsection{Results}\label{res.pg.i-c}
The major purpose to investigate the cylindrical configuration shown in \Rfi{Mvmsci.c} consists in emphasising the impact of the interface curvature on the thermal fields.
Accordingly, we have selected as baseline the computational case 1 in \Rta{cn.i}, relative to planar interface, whose results are shown in \Rfi{pg-c1T.i}: the solids are made by same material, their and the interface's stress response to thermal field is similar ($\beta_{\tsub{1}}=\beta_{\tsub{2}}=\tscn$), the flatness of the interface presupposes tension continuity [\REq{sbe.mom.cf.p.fin}] which, in combination with the additional equation [\REq{mfc+.pc.f}], implies all-temperature continuity [$\Tt(t) = T_{\tsub{1}}(\zerop,t) = T_{\tsub{2}}(\zerom,t)$]. 
At steady state, the temperature profile is a straight line and the interface is basically \textit{transparent}.
We have repeated the calculation for the cylindrical configuration with the same values of the characteristic numbers of case 1 in \Rta{cn.i} and with $ \tsct / \beta_{\tsub{1}} R = 0.1 $ and have found out that
\begin{figure}[h]
	\fbox{\includegraphics[bb = 40 65 595 532 , clip , width=.975\columnwidth]{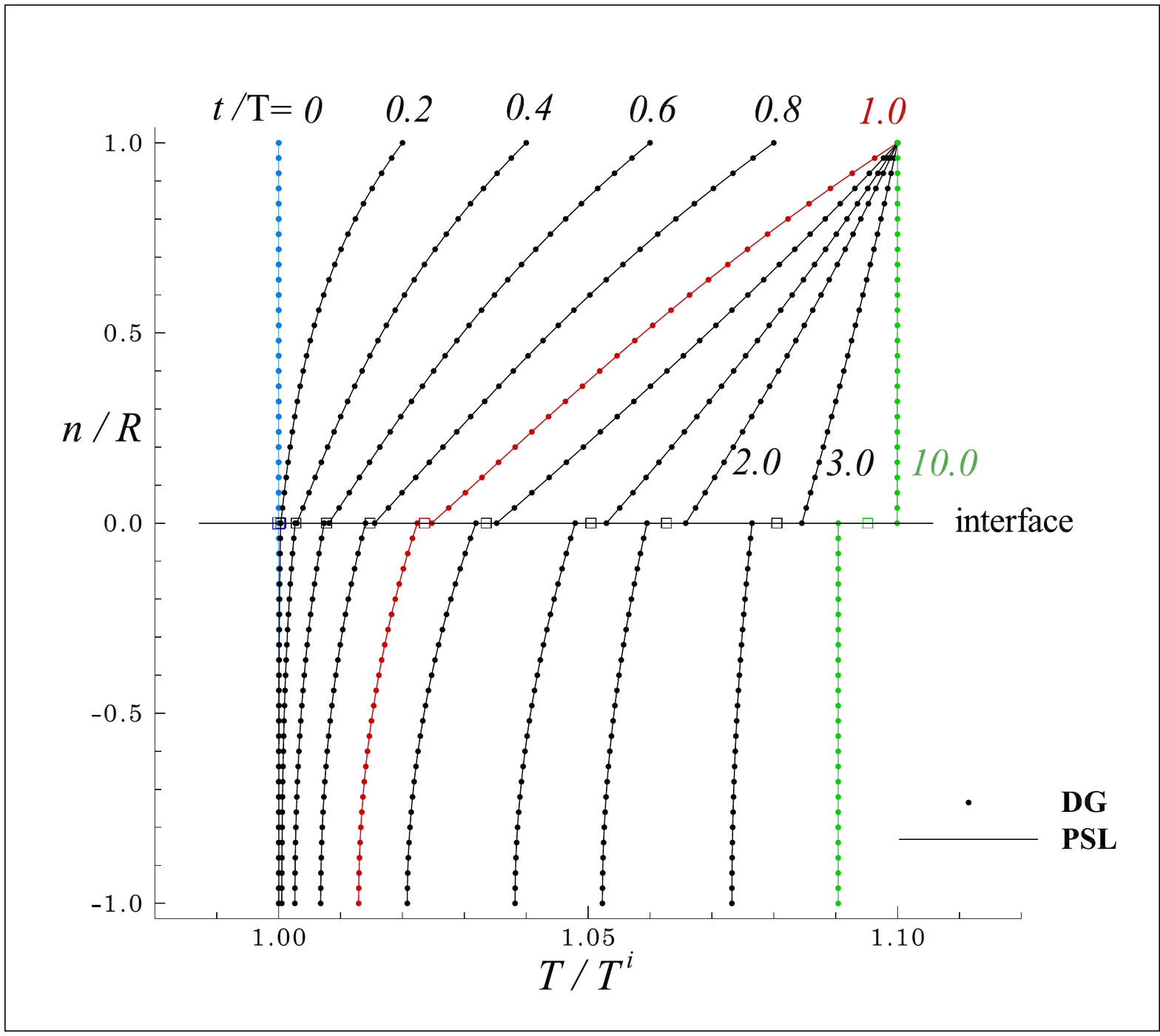}}
	\caption{Computational case 1 of \Rta{cn.i} for the cylindrical configuration of \Rfi{Mvmsci.c}. Curvature characteristic number $ \tsct / \beta_{\tsub{1}} R = 0.1 $.}
	\label{pg-c1T.i.c}
\end{figure}
the situation is different for a curved interface because the presence of the curvature breaks down tension continuity [\REq{sbe.mom.cf.c}] with the remarkable outcome that temperature continuity does not settle in even for solids made by same material.
This occurrence is clearly indicated by the temperature profiles shown in \Rfi{pg-c1T.i.c}. 
At steady state, the heat flux vanishes everywhere and the temperature profiles are uniform. 
The external solid reaches the final temperature (${T} / T^{\tsup{i}} = 1.1$) imposed on its external circumference ($n/R=1$) but the internal solid remains a bit cooler (${T} / T^{\tsup{i}} \simeq 1.09 $) because a fraction of the thermodynamic energy transported by diffusion during the transient is absorbed by the interface. 
The absorption process [\REq{sbe.te.cf.c.f}] is reflected in the interface-temperature (squares in \Rfi{pg-c1T.i.c}) growth, in this case as the mean value between the temperatures of the solids, with final settling at a steady-state value ${\Tt} / T^{\tsup{i}} \simeq 1.095 $.
The thermal protection exercised by the interface on the internal solid is evident and consistent with the one illustrated in \Rfi{pg-c5T.i} for the planar interface and for different materials. \\

\section{Conclusions} \label{concl}

The tentative application of the phenomenological theory of interfacial interactions to our \textit{demonstrator} heat-transfer test case has shed constructive light on several aspects.

In the absence of interfaces (\Rse{httc-std}), it has brought forth, and helped us to understand, the importance of tension continuity [\REq{mfc}] as physical boundary condition on the same footing of heat-flux continuity [\REq{hfc}]. 
In our test case, \REq{mfc} governs unequivocally the establishment of the temperature jump at the separation surface between the solids (\Rfi{pg-c3T}). 
With our minds prejudiced by the standardly enforced temperature-continuity boundary condition [\REq{Tc}], we received this result with a bit of justified hesitation and decided, therefore, to seek alternative and independent confirmation via {\color{black} MD} calculations.
These were performed by our collaborators in the technical university of Eindhoven and the details will be published soon in a dedicated article \cite{svn2016xxx}; here we only anticipate that the {\color{black} MD} results are in very good agreement with those produced by our calculations and, above all, confirm the temperature jump at the separation surface. 
{\color{black} The MD intervention proved decisive to remove doubts and hesitation and we certainly look forward to its, hopefully forthcoming, fundamental support to obtain the necessary thermophysical properties of the interfaces.} 
We are (now) convinced about the key role owned by the tension-continuity boundary condition and the physical significance of results generated by its imposition; we also believe and foresee that it will produce novel, unexpected, and far reaching consequences when enforced in fluid-dynamics contexts with fluids in contact with solids.

In the presence of interfaces (\Rse{httc-int}), the phe\-nom\-e\-no\-log\-ical-theory idea flourishes into a self-contained theoretical elaboration that, in our opinion, goes a long way in the direction of the ``closed theories which could a priori predict ... catalytic properties'' predicated by Kovalev and Kolesnikov \cite{vk2005fd}. 
All parameters called for in the elaboration have physical foundation and, remarkably, there is no place and no need for any heuristic empirical construct such as the ACs.
Of course, we have enforced simplifying assumptions about those parameters but exclusively for the sake of simplicity in numerical calculations, a reasonable justification, and certainly never imperiling, to the best of our understanding, their physical meaningfulness.
We believe that the quantitative results described and discussed in \Rsed{res.pg.i}{res.pg.i-c} are encouraging; they provide ulterior evidence of the phenomenological-theory predictive power and of how its physical parameters control the interface role as provider of thermal either protection or amplification. 
Therefore, we feel comfortable to conclude that they support convincingly our MTGSI research program and make worthwhile further investigation along its conceptual pathway.
It is true that, for the time being, our theoretical elaboration shares the important problem related to the correct closure of the surface balance equations, introduced in \Rse{deibc}, with predecessors in other application domains but, given the preliminary nature of our study, we have considered acceptable for our test case to overcome the problem as described in \Rse{dme} without going (yet) into deeper details. 
Indeed, we have not investigated the possible ramifications and implications of the surface-entropy balance equation which has already proven useful \textit{to fix} conceptual uncertainties \cite{db1976p,aj2014cmame,ln1979aa,ls2011rmp}.
We are obviously conscious of the importance of such a problem, are confident in the invaluable help harvestable from in-depth familiarisation with the pertinent literature and are prepared to confront it head-on in our future work of more fluid-dynamics nature
that will concentrate on the MTGSI-idea application to a two-dimensional flow past a cylinder with and without interface, as sketched in \Rfi{Mvmsci.c.fd}.


%
%

\begin{acknowledgements}
One of the authors (DG) wishes to dedicate his contribution in this work to his former student Birte Schmidtmann \cite{bs2013vki} whose efforts should have been better recognised instead of being received with undeserved unscientific mistreatment when she presented physically consistent and correct results, although limited by simple modeling, that have been confirmed in full by the more complete physical model described in this work. The contributions of P.  Solano-L\'{opez} and J. M. Donoso have been supported by the ESA TRP contract no. 4000112582/14/NL/PA titled ``Novel methodology for gas-surface interactions in hypersonic reentry flows''.
\end{acknowledgements}

\bibliographystyle{spmpsci}      
\bibliography{mybibreflibrary}   

%
%

\end{document}